\documentclass[review]{elsarticle}
\usepackage{tabularx}
\usepackage{amsmath}
\usepackage{adjustbox,lipsum}
\extrafloats{2}
\journal{Journal of Advances in High Energy Physics}

\begin{document}
\begin{frontmatter}
\title{Charged-particle multiplicity moments as described by shifted Gompertz distribution in $e^{+}e^{-}$, $p\overline{p}$ and $pp$ collisions at high energies}

\author{Aayushi Singla}
\ead{aayushi.singla@cern.ch}
\author{M.Kaur\fnref{myfootnote}}
\ead{manjit@pu.ac.in}
\address{Physics Department, Panjab University, Chandigarh, India-160014}

\begin{abstract}
In continuation of our earlier work, in which we analysed the charged particle multiplicities in leptonic and hadronic interactions at different center-of-mass energies in full phase space as well as in restricted phase space using the shifted Gompertz distribution, a detailed analysis of the normalized moments and normalized factorial moments is reported here.~A two-component model in which a probability distribution function is obtained from the superposition of two shifted Gompertz distributions, as introduced in our earlier work, has also been used for the analysis.~This is the first analysis of the moments with the shifted Gompertz distribution.~Analysis has also been performed to predict the moments of multiplicity distribution for the $e^{+}e^{-}$ collisions at $\sqrt s$ = 500~GeV at a future collider.  
\end{abstract}

\begin{keyword}
Charged multiplicities, Probability Distribution Functions, Factorial moments
\end{keyword}

\end{frontmatter}

\section{Introduction}
In one of our recent papers, we introduced a statistical distribution, the shifted Gompertz distribution to investigate the multiplicity distributions of charged particles produced in $e^{+}e^{-}$ collisions at the LEP, $p\overline{p}$ interactions at the SPS and $pp$ collisions  at the LHC at different center of mass energies in full phase space as well as in  restricted phase space \cite{shGomp}.
~A distribution of the largest of two independent random variables, the shifted Gompertz distribution was introduced by Bemmaor \cite{Bema} as a model of adoption of innovations.~One of the parameters has an exponential distribution and the other has a Gumbel distribution, also known as log-Weibull distribution.~The non-negative fit parameters define the scale and shape of the distribution.~Subsequently, the shifted Gompertz distribution has been widely studied in various contexts \cite{Jon, Jod, Jim}.~In our earlier work \cite{shGomp} by studying the charged particle multiplicities, we showed that this distribution can be successfully used to study the statistical phenomena in high energy $e^{+}e^{-}$, $p\overline{p}$  and $pp$ collisions at the LEP, SPS and LHC colliders, respectively.

A multiplicity distribution is represented by the probabilities of n-particle events as well as by its moments or its generating function.~The aim of the present work is to extend the analysis by calculating the higher moments of a multiplicity distribution.~Because the moments are calculated as derivatives of the generating function, the moment analysis is a powerful tool which helps to unfold the characteristics of multiplicity distribution.~The multi-particle correlations can be studied through the normalized moments and normalized factorial moments of the distribution.~The dependence of moments on energy can also reveal the KNO (Koba, Nielsen and Olesen) scaling \cite{KNO, UA51, UA52} conservation or violation.~Several analyses of moments have been done at different energies, using different probability distribution functions and different types of particles \cite{Ash, Mic, Capel, Nao}.~The higher moments also can identify the correlations amongst produced particles. 

In section~2, formulae for the Probability Distribution Function (PDF) of the shifted Gompertz distribution, normalized moments and the normalized factorial moments used for the analysis are given.~A two-component model has been used and modification of distributions carried out, in terms of these two components; one from soft events and another from semi-hard events.~Superposition of distributions from these two components, by using appropriate weights is done to build the full multiplicity distribution.~When multiplicity distribution is fitted with the weighted superposition of two shifted Gompertz distributions, we find that the agreement between data and the model improves considerably.~The details of these fits are published in \cite{shGomp}.~The distributions have been fitted both in full phase space as well as in restricted rapidity windows for $p\overline{p}$ and $pp$ data and in five rapidity windows and in full phase space only for $e^+e^-$, in terms of soft and semi-hard components.

Section~3 presents the evaluations of moments from experimental data, the fitted shifted Gompertz distributions  and the fitted modified shifted Gompertz distributions.~Section~4 details the method of estimating uncertainties on the moments.~Discussion and conclusion are presented in Section~5.

\section{Shifted Gompertz distribution and Moments}

The particle production dynamics can be understood by analysing the charged particle multiplicity distribution as its measurements can provide relevant constraints for particle-production models.~Charged particle multiplicity is defined as the average number of charged particles $n$, produced in a collision at a given energy in the center of mass system.

In addition, the analysis of moments of the distribution is often used to study the patterns and correlations in the multi-particle final state of high-energy collisions in the presence of statistical fluctuations.~The fractal structures present in the multiplicity distributions have often been studied to search for the embedded constraints on the underlying particle production mechanism \cite{Fermi1, Fermi2}.~The observation of fractal structures is of great interest because it imposes strong constraints on the underlying particle-production mechanism.~We define  different kinds of moments as follows.

Let X be any non-negative random variable having the shifted Gompertz distribution with parameters b and $\beta$, where b $>$ 0 is a scale parameter and $\beta >$ 0 is a shape parameter.~The probability distribution function (PDF) of X is given by
\begin{equation}
P_X(x;b,\beta) = b e^{-(bx + \beta e^{-bx})}\big(1+\beta(1-e^{-bx})\big), \>\>\>\>where\>\>\>x>0
\end{equation}

The raw moments ($c_n$) and factorial moments ($f_n$) are defined as:
\begin{equation}
c_n = E[X^{n}] \>\>\>\>and\>\>\>\> f_n = E[(X)(X-1)(X-2)....(X-(n-1))]
\end{equation}

Whereas, the normalized moments ($C_n$) and normalized factorial moments ($F_n$) are defined as following:
\begin{equation}
C_n = \frac{E[X^{n}]}{(E[X])^{n}} \>\>\>\>and\>\>\>\> F_n = \frac{E[(X)(X-1)(X-2)....(X-(n-1))]}{(E[X])^{n}}
\end{equation} 

$n$ as a natural number ranging from 1 to $\infty$.~The Mean value (E[X]) of Shifted Gompertz distribution is given by 
\begin{equation}
E[X] = \frac{1}{b}( \gamma + \ln\beta + \frac{1 - e^{-\beta}}{\beta} + \Gamma[0,\beta] )
\end{equation}

and the c$_2$ moment is given by
\begin{equation}
c_2 = E[X^{2}] = \frac{2}{b^{2} \beta} \Big(\gamma + \Gamma[0,\beta] + \beta^{2}\>_3F_3[\{1,1,1\},\{2,2,2\},-\beta] + \ln\beta\Big) 
\end{equation}

The higher order raw moments ($c_n$) can be found by the Moment Generating Function of the Shifted Gompertz distribution[[$b$,$\beta$],t] 
\begin{equation}
e^{-\beta} - (1 + \frac{t}{b\beta}) \beta^{\frac{t}{b}}(\Gamma[1-\frac{t}{b}] - \Gamma[1-\frac{t}{b},\beta])
\end{equation}

Also $f_2$ Moment is given by 
\begin{equation}
\begin{aligned}
f_2 = E[(X)(X-1)] ={}&\frac{2}{b^{2}\beta}\Big(\gamma + \Gamma[0,\beta] + \beta^{2} \>_3F_3[\{1,1,1\},\{2,2,2\},-\beta] + \ln\beta\Big) - \\
&\frac{\Big(1-e^{-\beta} + \beta (\gamma + \Gamma[0,\beta] + \ln\beta)\Big)}{b \beta}
\end{aligned}
\end{equation}

The higher factorial moments ($f_n$) can be found by the Generating Function of Shifted Gompertz Distribution[[$b$,$\beta$],t] 
\begin{equation}
e^{-\beta} - \beta^{\frac{\ln t}{b}} (\Gamma[1-\frac{\ln t}{b}] - \Gamma[1-\frac{\ln t}{b},\beta]) (1  + \frac{\ln t}{b \beta})
\end{equation}
where\\
(i) $\gamma$ $\approx$ 0.5772156 stands for the Euler constant (also referred to as Euler-Mascheroni constant).
\\
(ii) $\Gamma$[s] the Euler Gamma function and $\Gamma[s,x]$ the incomplete Gamma function defined below;\\
\begin{equation}
\Gamma[s] = \int_{0}^{\infty} t^{s-1} e^{-t}dt \hspace{ 1cm} \Gamma[s,x] = \int_{x}^{\infty}  t^{s-1} e^{-t}dt
\end{equation}
\\
(iv) $_3F_3$ is a Generalized Hypergeometric function.\\ 
\begin{equation}
_3F_3[\{1,1,1\},\{2,2,2\},-\beta] = \sum_{k=1}^{\infty} \frac{(-1)^{k+1} \beta^{k+1}}{k! k^{2}} 
\end{equation}

\subsection{ Modified Shifted Gompertz Distribution}

It is well established that at high energies, charged particle multiplicity distribution in full phase space becomes broader than a Poisson distribution.~The most widely adopted, Negative Binomial distribution \cite{NBD} to describe the multiplicity spectra, fails to explain the experimental data.~As a corrective measure to explain the failure, a two-component approach, was introduced by A. Giovannini et al \cite{NBD}.~The details are included in our earlier publication  \cite{shGomp} on the Shifted Gompertz distribution.
 
To better explain the data at high energies, a superposition of two shifted Gompertz distributions, which are interpreted as soft and hard components, is used.~The multiplicity distribution is produced by adding weighted superposition of multiplicity in soft events and multiplicity distribution in semi-hard events.~This approach combines merely two classes of events and not two different particle-production mechanisms.~Therefore, no interference terms need to be introduced.~The final distribution is the superposition of the two independent distributions.~We call it 'modified shifted Gompertz distribution'.

\small
\begin{equation}
P(n)=\alpha P_{soft}^{shGomp}(n)+(1-\alpha)P_{semi-hard}^{shGomp}(n)
\end{equation} 
\normalsize
Adopting this approach  for the multiplicity distributions in $e^+e^-$, $pp$ and $p\overline{p}$ collisions at high energies, the data at different energies are fitted with the distribution which involves five parameters as given below;
\small
\begin{equation}
P_{n}(\alpha:b_1,\beta_1;b_2,\beta_2)=
\alpha P_{n}(soft) + 
(1-\alpha)P_{n}(semi\textrm{-}hard) 
\end{equation} 
where $\alpha$ is the fraction of soft events, ($b_1$, $\beta_1$) and ($b_2$, $\beta_2$) are respectively the scale and shape parameters of the two distributions.

Values of $\alpha$ evaluated for different interactions, were published in the tables in our previous paper \cite{shGomp}.~In figure 2, we show the variation of $\alpha$ as a function of collision energy and pseudorapidity for $pp$ interactions at LHC energies.~The values show that the alpha decreases with collision energy as well as with increasing pseudo-rapidity window.

\normalsize

\section{Analysis and Results}
Calculations of the normalized moments and the normalized factorial moments are presented by using the data from different experiments and following three collision types;\\
i) $e^{+}e^{-}$ annihilations at different collision energies, from 91 GeV up to the highest energy of 206 GeV at LEP2, from two experiments L3 \cite{L3} and OPAL \cite{OPAL91, OPAL1, OPAL2,OPAL3} are analysed.\\
ii) $pp$ collisions at LHC energies from 900 GeV, 2360 GeV and 7000 GeV \cite{CMS} are analysed for five intervals of increasing extent in pseudorapidity $|\eta| <$ 0.5 up to $|\eta| <$ 2.4.\\
iii) $p\overline{p}$ collisions at energies from  200 GeV, 540 GeV and 900 GeV \cite{UA51,UA52} are analysed in full phase space as well as in pseudorapidity intervals from $|\eta|<$ 0.5 up to $|\eta|<$5.0, 
where $\eta$ is defined as $-ln[tan(\theta/2)]$, and $\theta$ is the polar angle of the particle with respect to
the counter-clockwise beam direction.\\
The PDF defined by equation~(1) is used to fit the experimental data on charged particle multiplicity distributions, for the shifted Gompertz function and the modified (two-component) function.~Results from these fits to the above mentioned data were published in our earlier work \cite{shGomp}.~As an example, results for L3 data are shown in figure~1.~To avoid repetition, the details of other figures are not given here.~It was shown that the data are very well explained by the modified shifted Gompertz distribution and the $\chi^{2}$ values for the fits in almost all cases reduce substantially.~In the present analysis we calculate the normalized moments and the normalized factorial moments defined in equations (2-8).

\begin{figure}[ht]
\includegraphics[width=4.8 in, height =2.8 in]{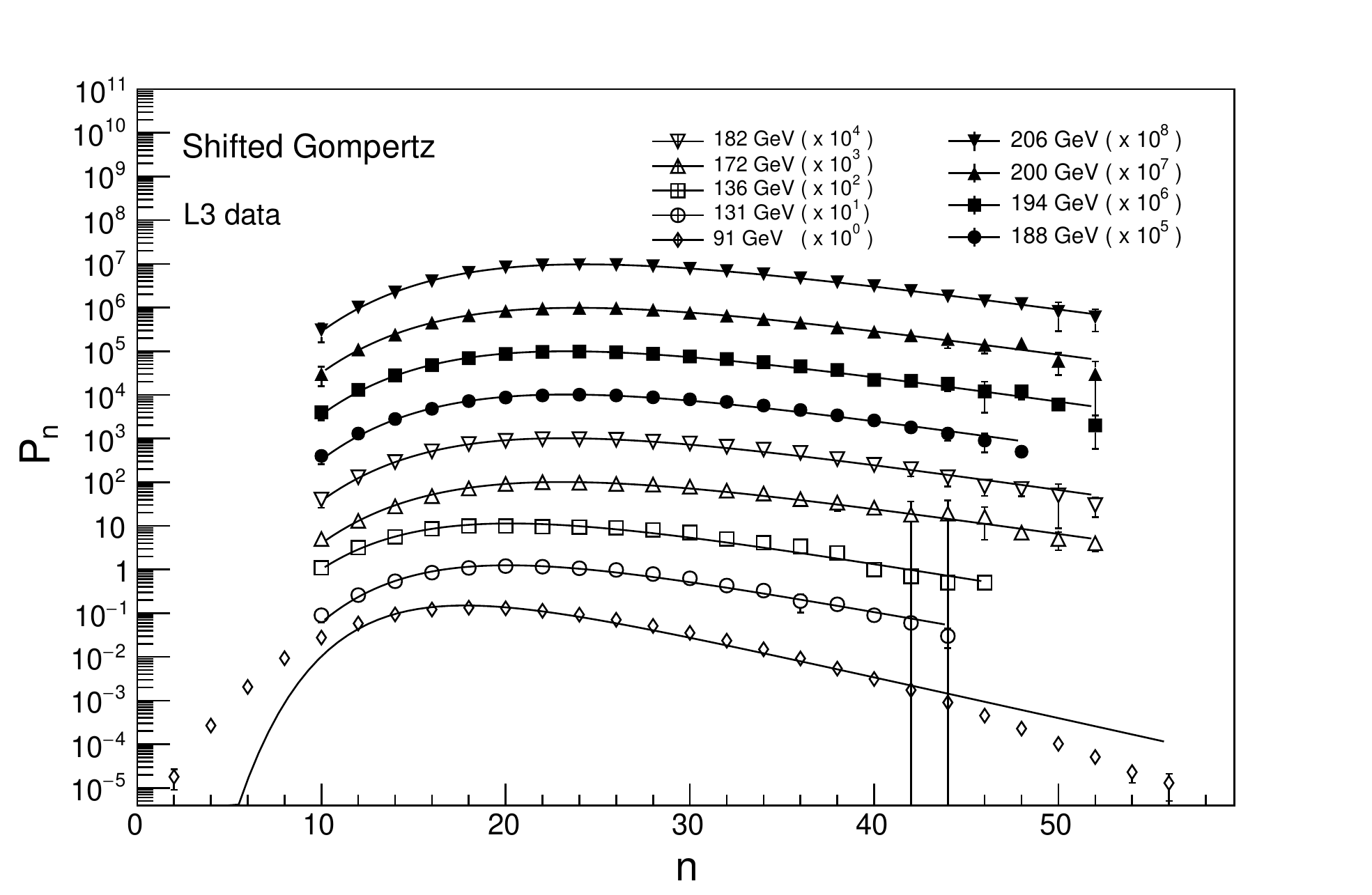}
\includegraphics[width=4.8 in, height =2.8 in]{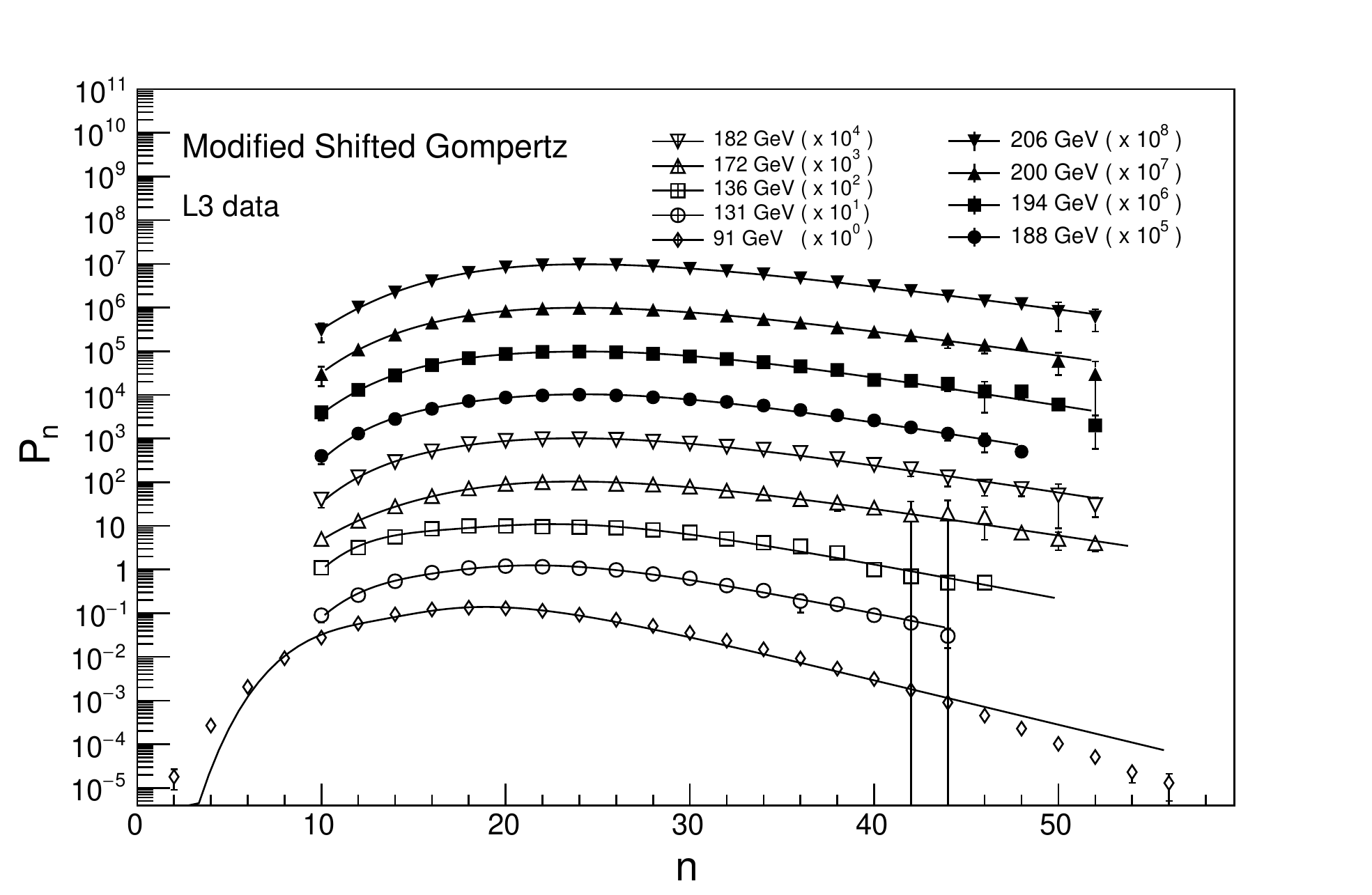}
\caption{Charged multiplicity distributions from L3 experiment.~Solid lines represent the shifted Gompertz distribution and the modified shifted Gompertz distributions}
\end{figure}

Figures~3-6 show the normalized moments ($C_q$) and the normalized factorial moments ($F_q$) calculated from the data of two experiments at LEP, at different energies for $e^{+}e^{-}$ collisions and also from the fitted shifted Gompertz and modified shifted Gompertz distributions.~The values of the moments are documented in table~1. 

Figures~7-10 show the normalized moments ($C_q$) and the normalized factorial moments ($F_q$) calculated from the data and from the modified shifted Gompertz distribution for the $p\overline{p}$ data at energies from 200, 540 and 900 GeV, in four rapidity bins and full phase space.~The values of the moments are documented in table~2. 

Figures~11-14 show the normalized moments ($C_q$) and the normalized factorial moments ($F_q$) calculated from the data and  from the modified shifted Gompertz distribution for the $pp$ data at energies from 900, 2360 and 7000 GeV in five rapidity bins.~The values of the moments are documented in table~3.

It is observed that the moments decrease with the increase in rapidity window, at all energies.~It is also interesting to compare the results from $pp$ and $p\overline{p}$ collisions at the same c.m. energy.~From the comparison at 900 GeV, we find that the moments have higher values in case of $p\overline{p}$ than $pp$.~For example, we include figure~15 to show the dependence for $|\eta|<$ 0.5. ~However values of multiplicity for $pp$ collisions at 900 GeV in full phase space are not available.~In addition, for $pp$ and for $p\overline{p}$ the rapidity intervals are also different.~So a direct comparison is not feasible. 

Figures~5-14, described above, are shown only for the modified shifted Gompertz distributions.~To avoid repetition and cluttering of figures, the figures for shifted Gompertz distributions are only included for $e^{+}e^{-}$ collisions and are not included for pp and $p\overline{p}$ data.~However the values of the moments are given in the tables~1-3.  

The predictions for normalized moments and normalized factorial moments are also made for $e^+e^−$ collisions at 500 GeV at a future collider.~By using the shifted Gompertz distribution, the prediction for probability distribution is made, as shown in figure~16.~Using this predicted distribution, 1$\sigma$ confidence interval band, moments have been calculated, as given in table~4.

It is observed that in case of $p\overline{p}$ and $pp$ interactions, $C_2$, $C_3$ and $F_2$, $F_3$ remain roughly constant with energy while higher moments $C_4$, $C_5$, $F_4$, $F_5$ show an increase with increasing energy.~The increase becomes more evident for larger rapidity windows.~This leads to the depiction of violation of KNO scaling at high energies.~Same conclusions have been reported in the reference \cite{CMS} for $pp$ collisions at the LHC.~However for the $e^+e^-$ collisions, the moments are roughly independent of energy.~This is expected as the collision energy is low, nearly at the onset of energy range, which marks the start of KNO scaling violation.

\begin{figure}[ht] 
\includegraphics[width=4.8 in, height =2.8 in]{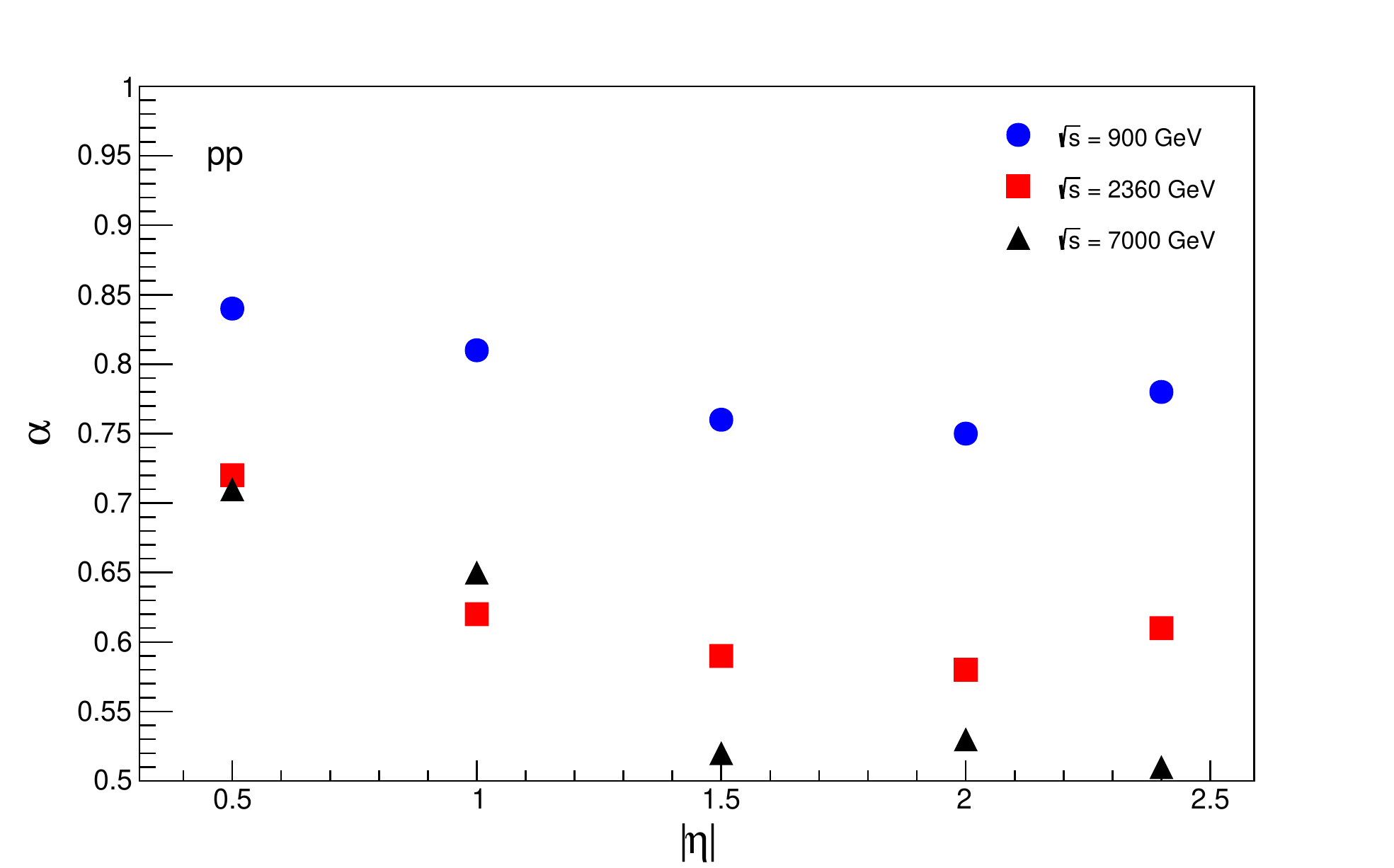}
\caption{Dependence of $\alpha$ on c.m. energy and pseudprapidity for pp interaction}
\end{figure}

\begin{figure}[ht]
\includegraphics[width=4.8 in, height =2.8 in]{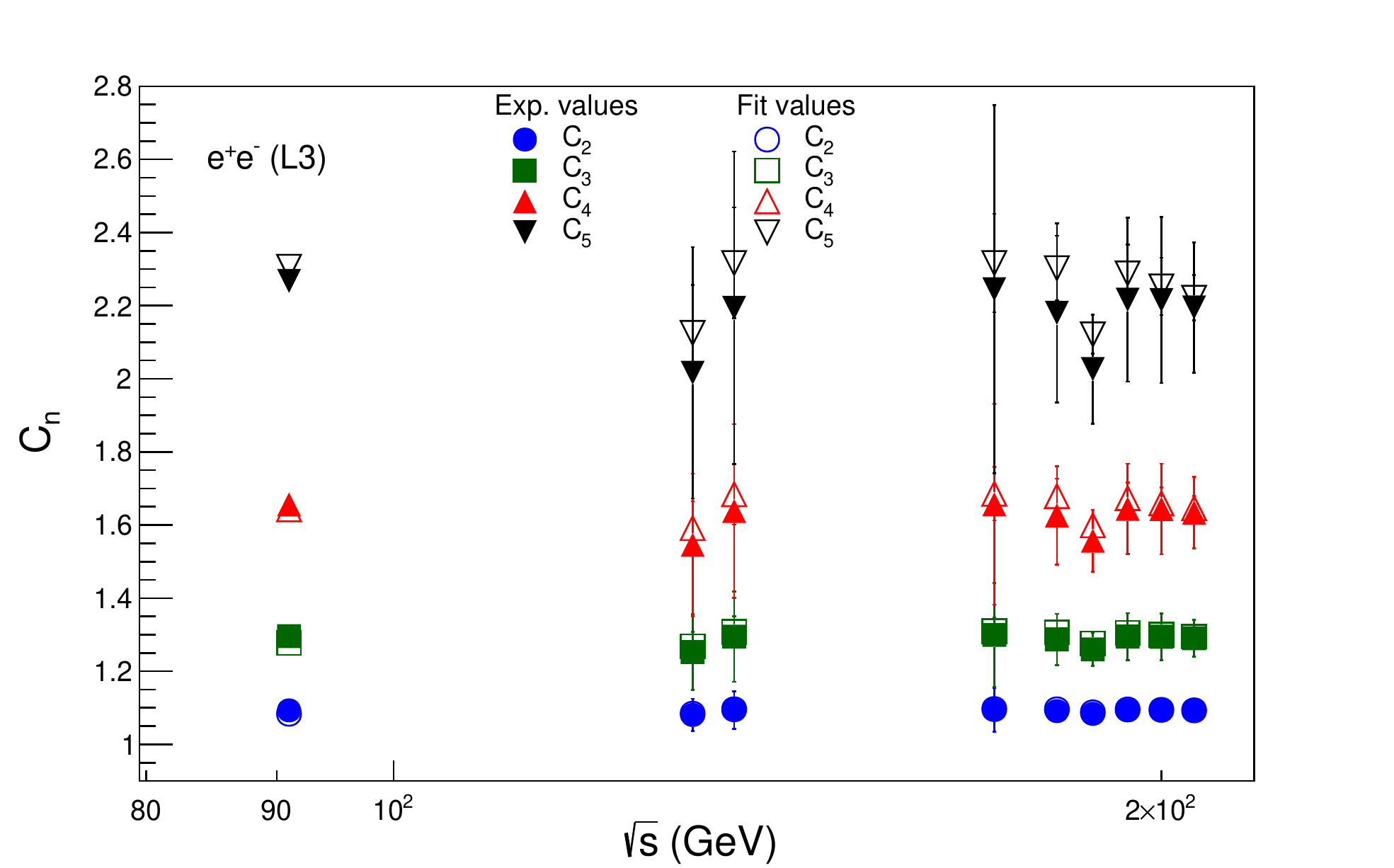}
\includegraphics[width=4.8 in, height =2.8 in]{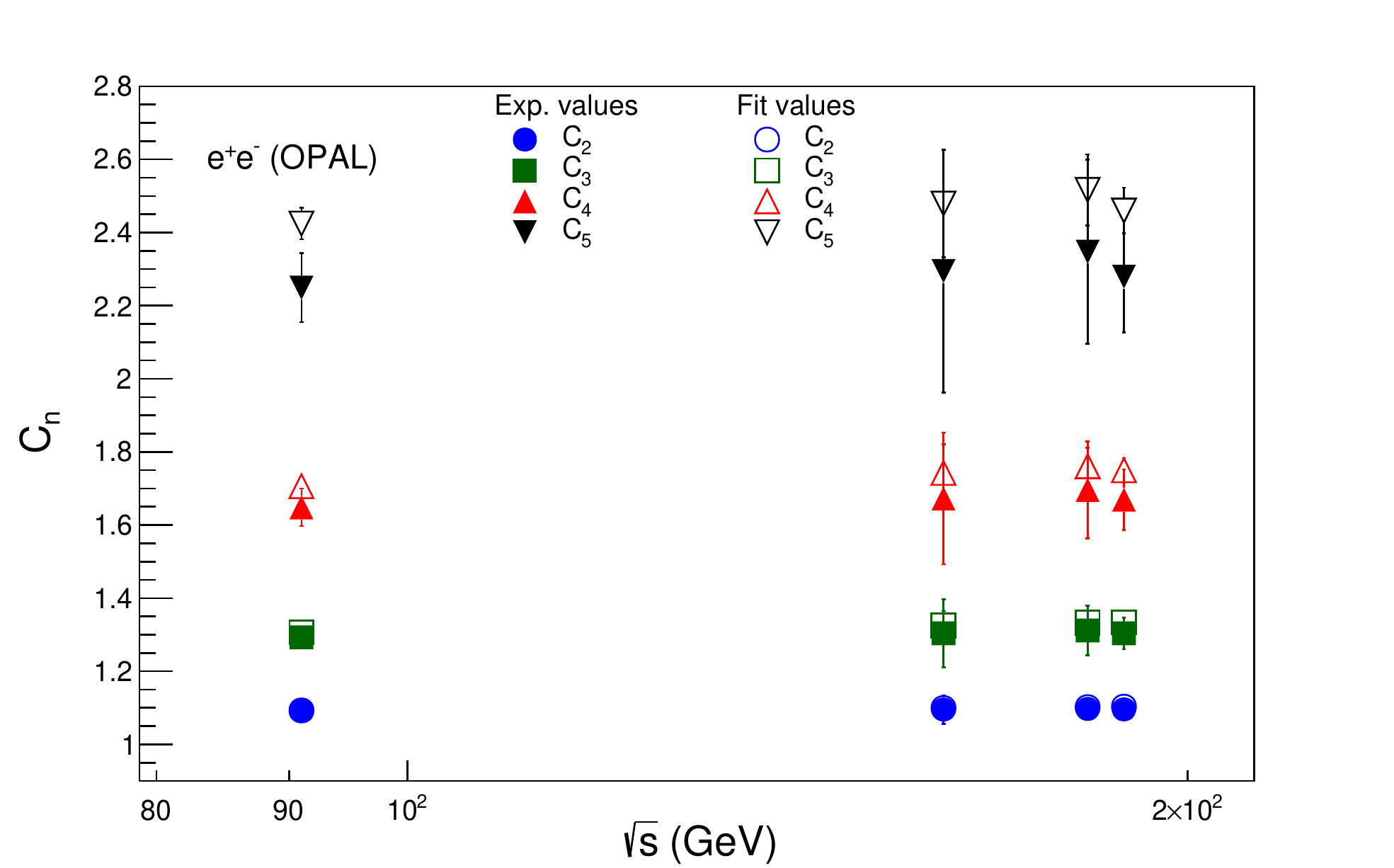}
\caption{Normalized moments of shifted Gompertz distribution for the $e^{+}e^{-}$ collisions recorded by the L3 and OPAL experiments at different energies.}
\end{figure}

\begin{figure}[ht]
\includegraphics[width=4.8 in, height =2.8 in]{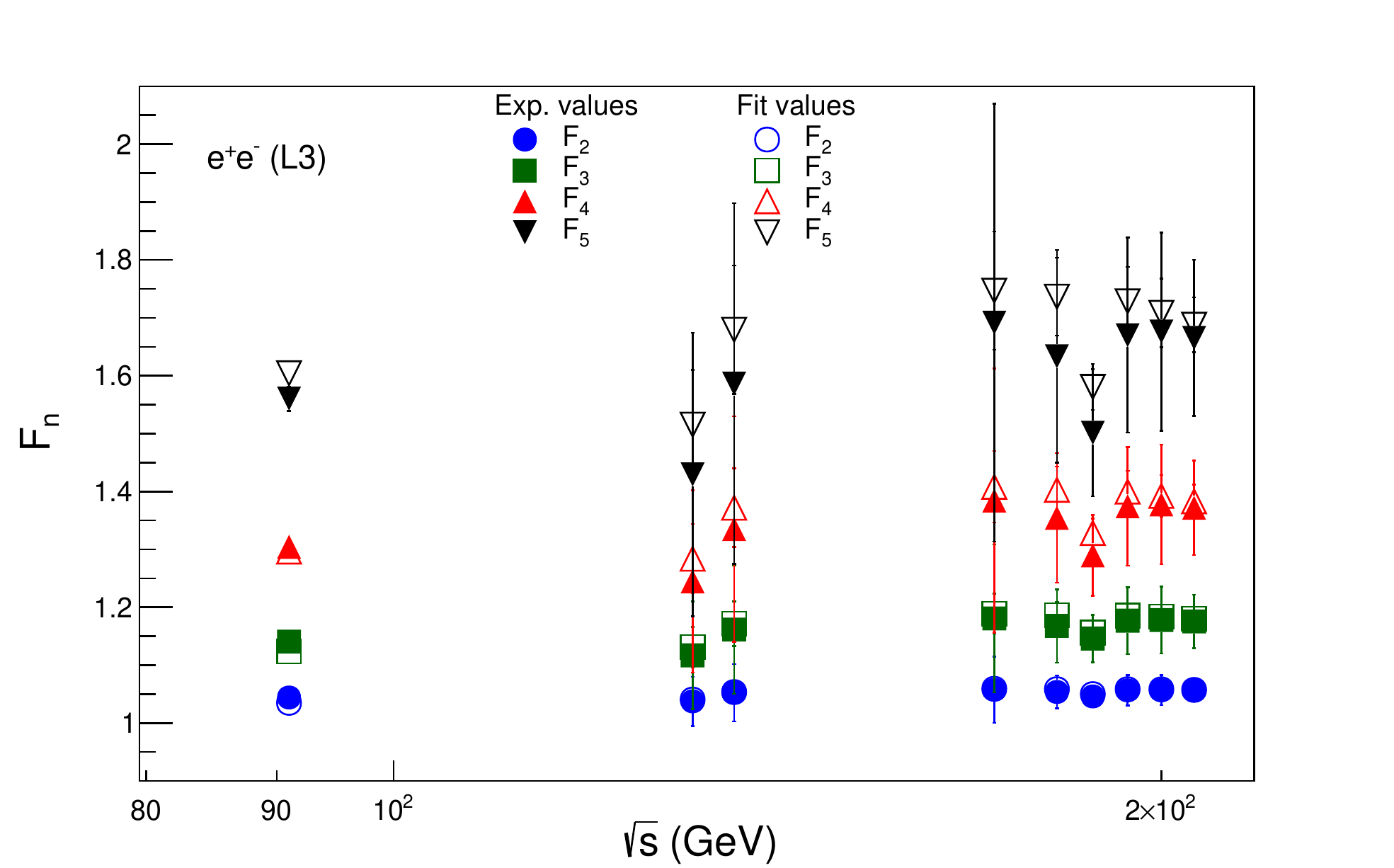}
\includegraphics[width=4.8 in, height =2.8 in]{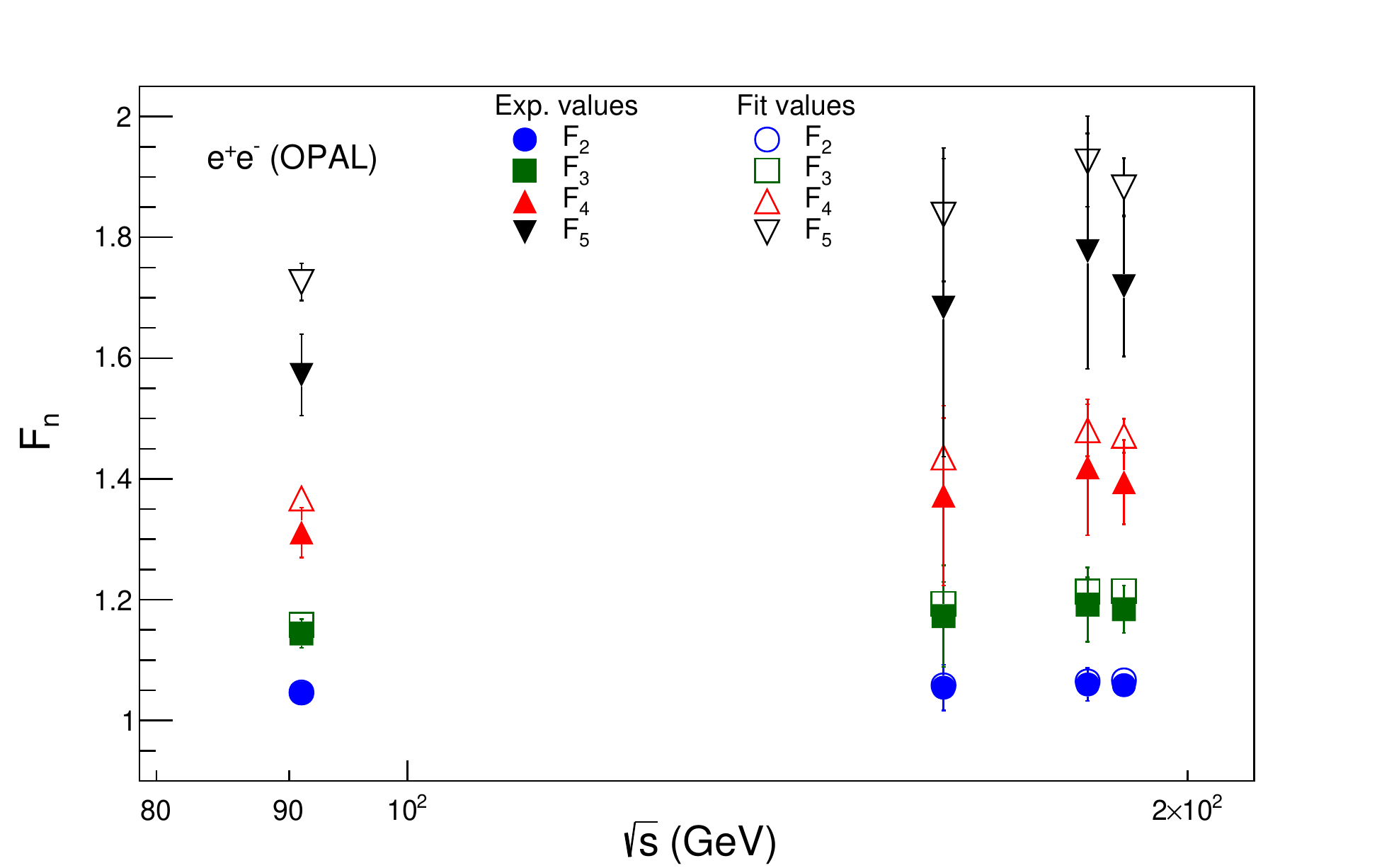}
\caption{Normalized factorial moments of shifted Gompertz distribution for the $e^{+}e^{-}$ collisions recorded by the L3 and OPAL experiments at different energies.}
\end{figure}

\begin{figure}[ht]
\includegraphics[width=4.8 in, height =2.8 in]{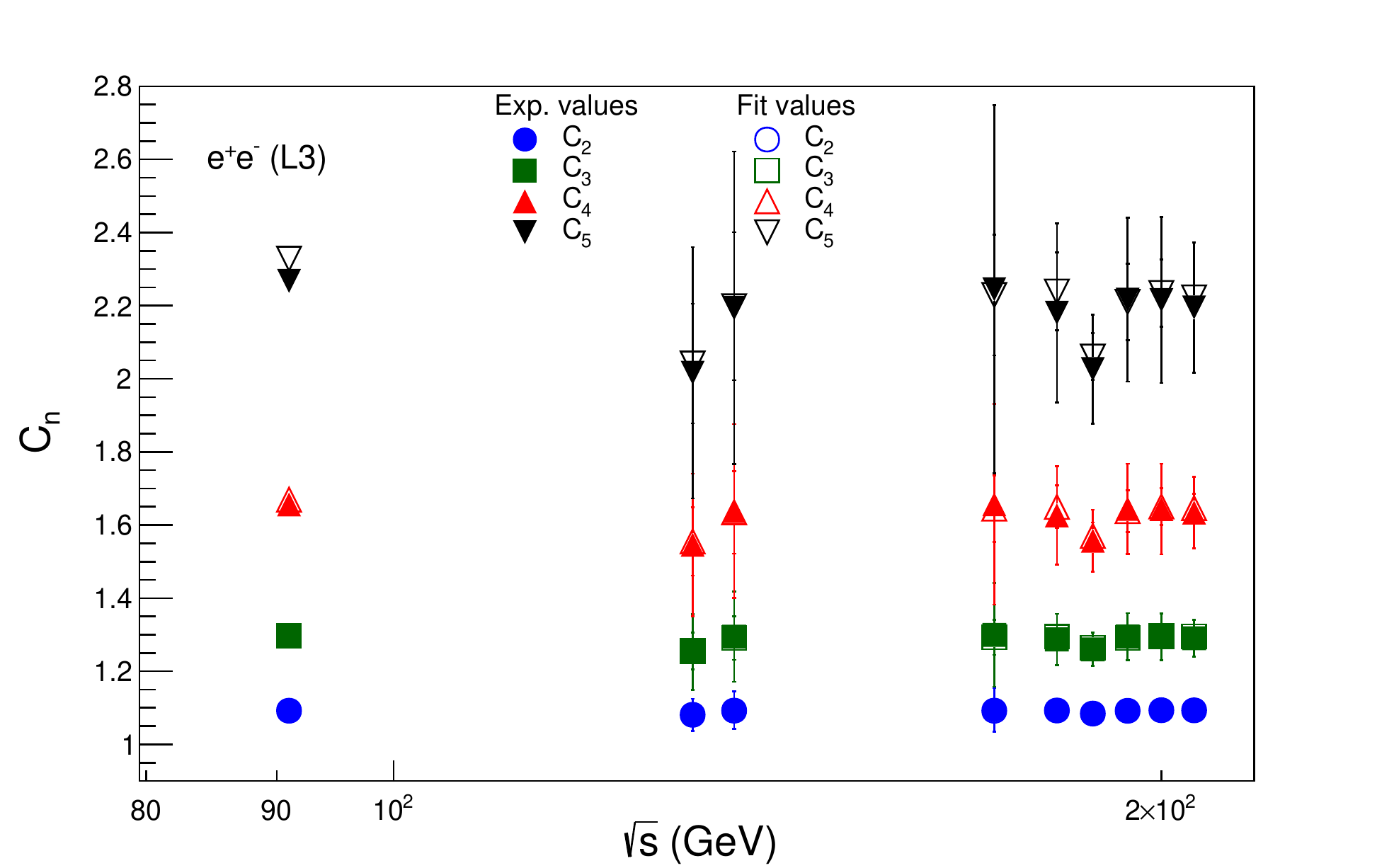}
\includegraphics[width=4.8 in, height =2.8 in]{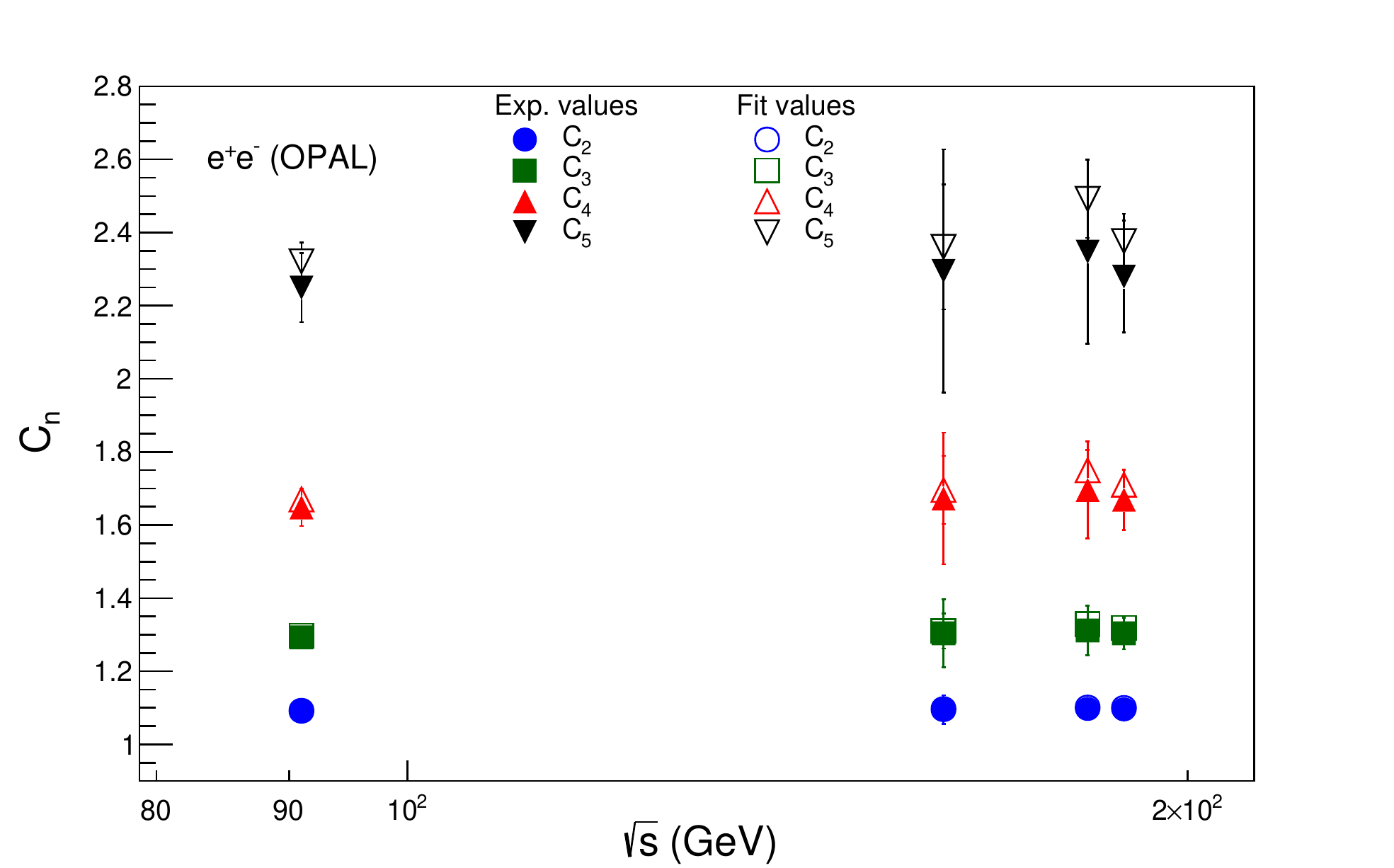}
\caption{Normalized moments of modified shifted Gompertz distribution for the $e^{+}e^{-}$ collisions recorded by the L3 and OPAL experiments at different energies.}
\end{figure}

\begin{figure}[ht]
\includegraphics[width=4.8 in, height =2.8 in]{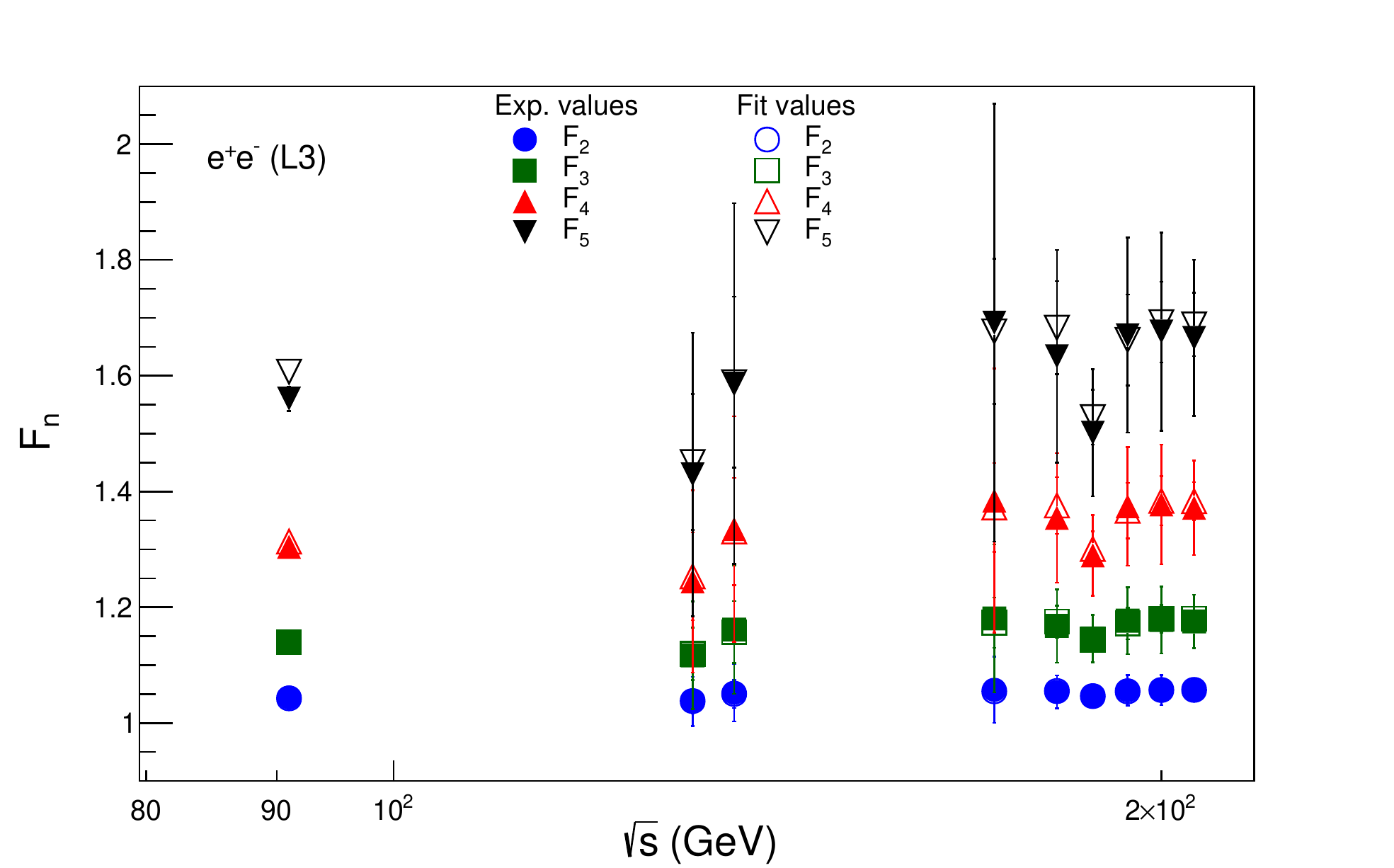}
\includegraphics[width=4.8 in, height =2.8 in]{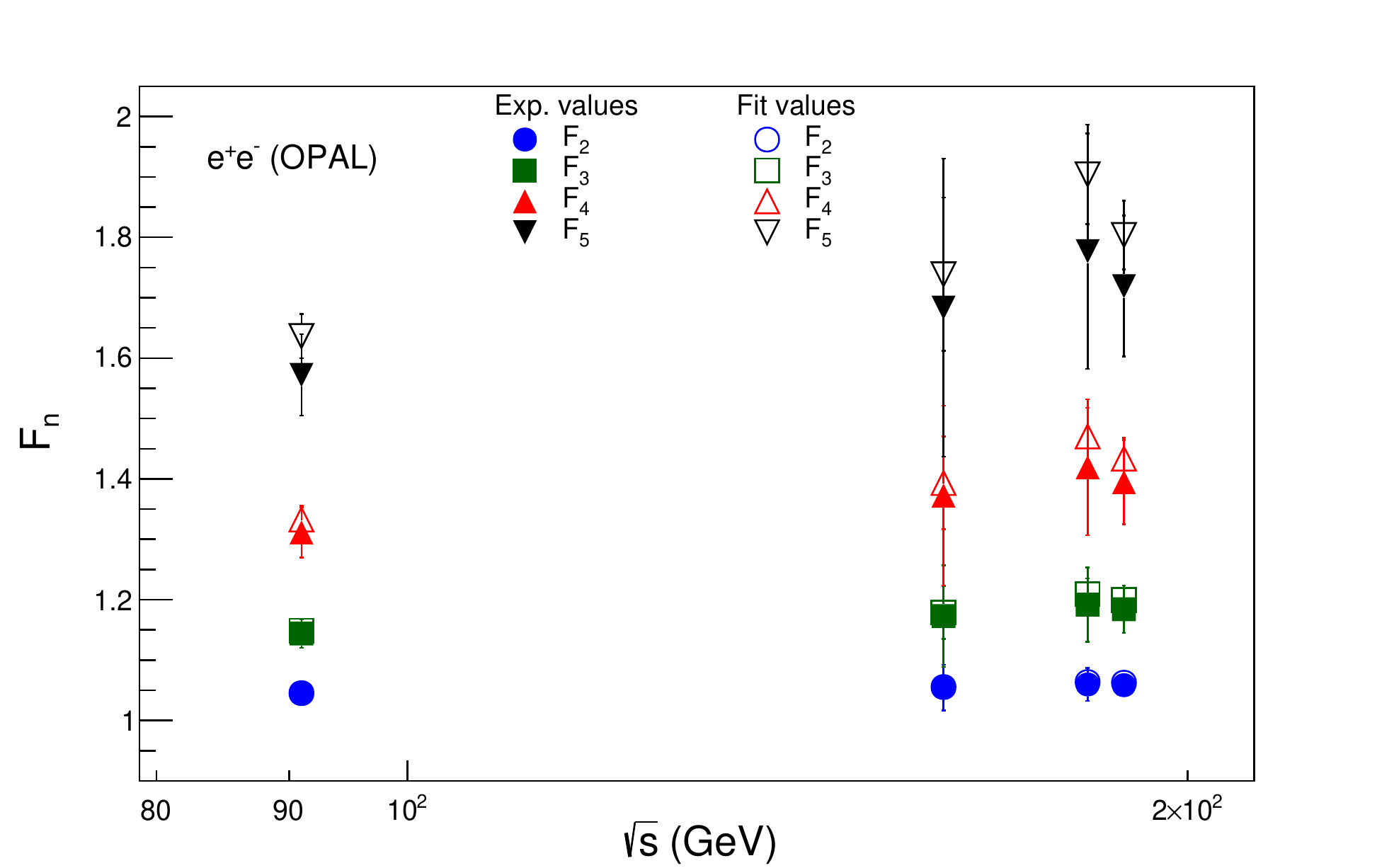}
\caption{Normalized factorial moments of modified shifted Gompertz distribution for the $e^{+}e^{-}$ collisions recorded by the L3 and OPAL experiments at different energies.}
\end{figure}
\begin{figure}[ht]
\includegraphics[width=4.8 in, height =2.6 in]{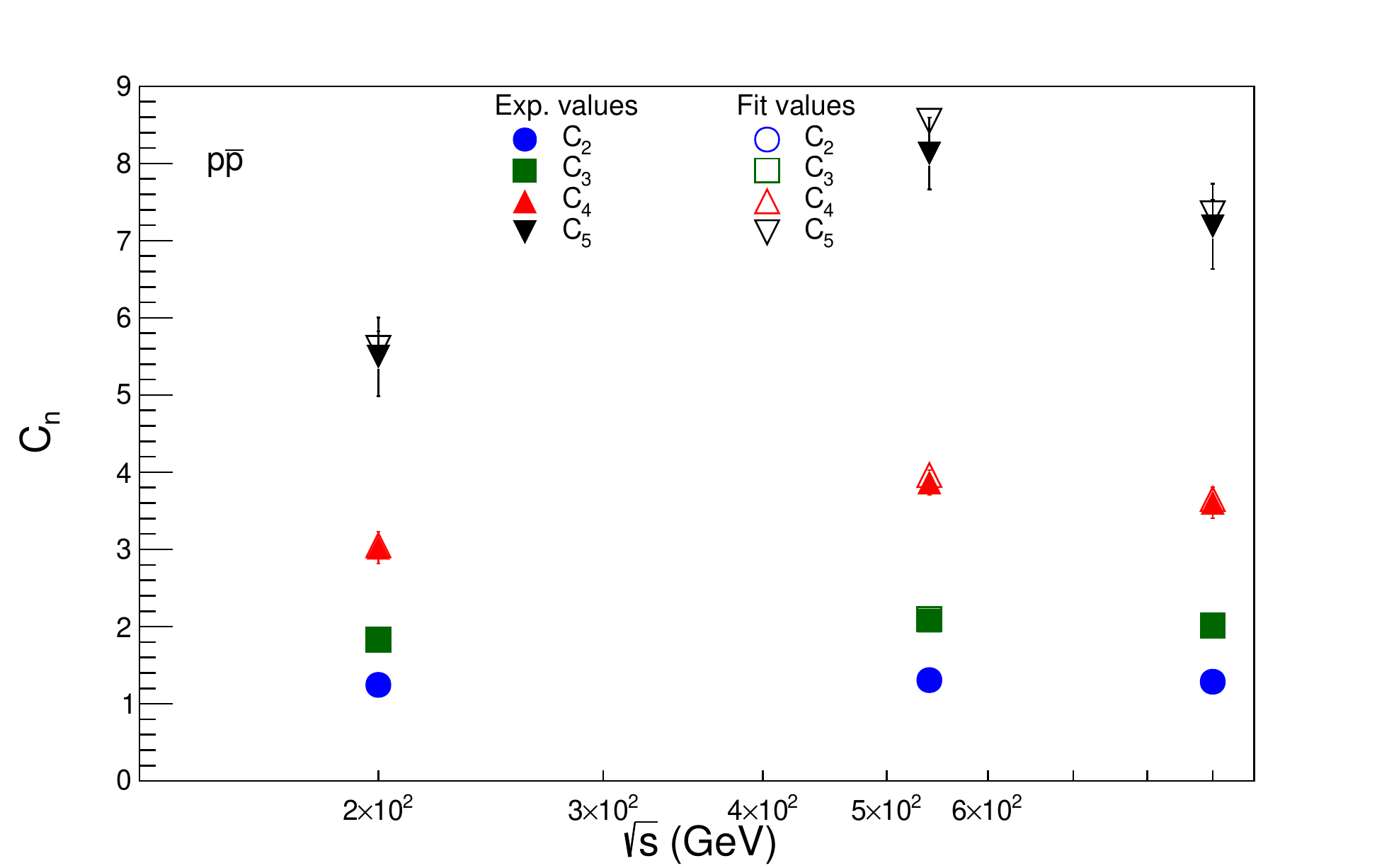}
\includegraphics[width=4.8 in, height =2.6 in]{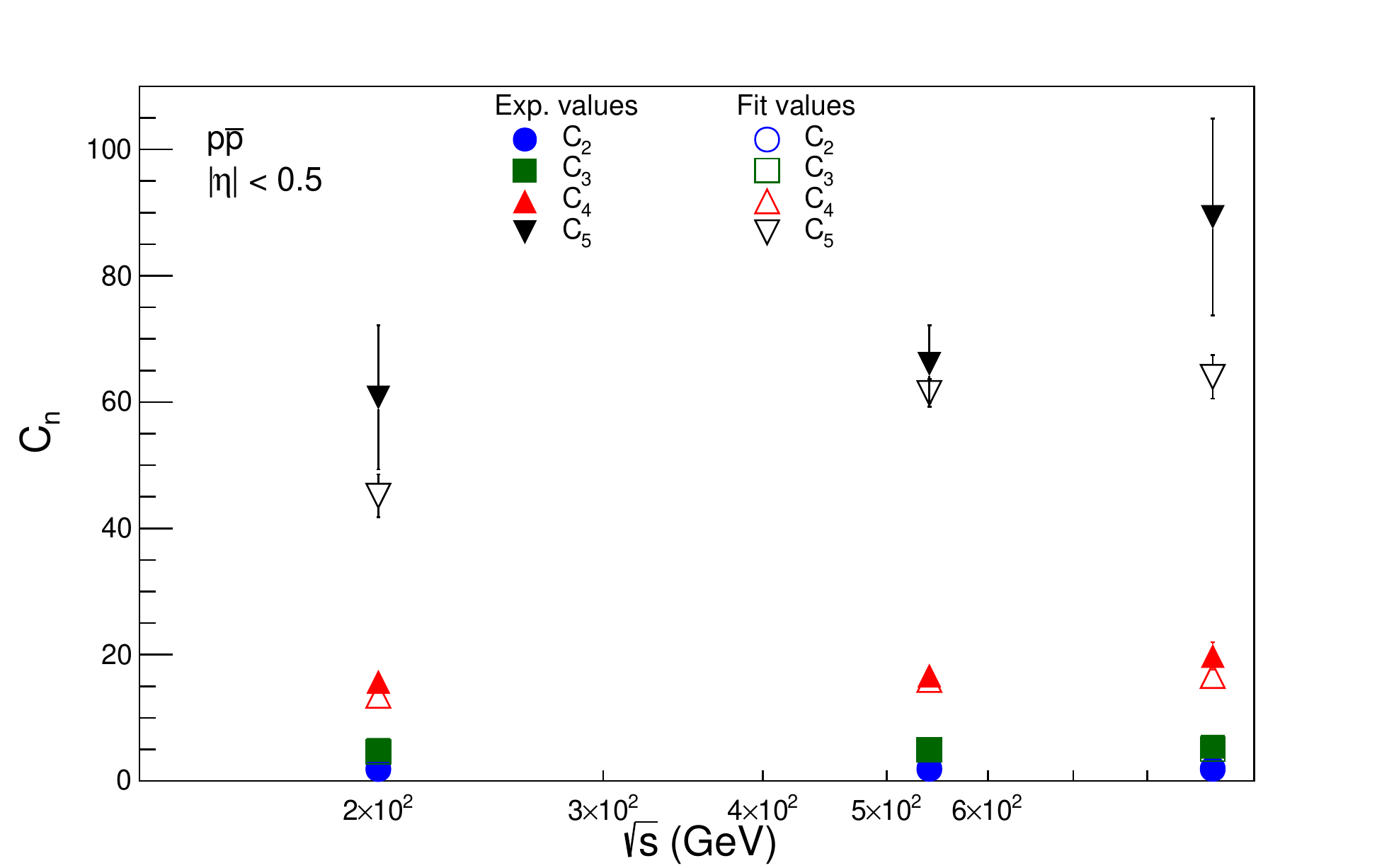}
\caption{Normalized factorial moments of modified shifted Gompertz distribution for the  $p\overline{p}$ collisions recorded by the $UA5$ experiment at different energies and in different rapidity bins.}
\end{figure}
\begin{figure}[ht]
\includegraphics[width=4.8 in, height =2.43 in]{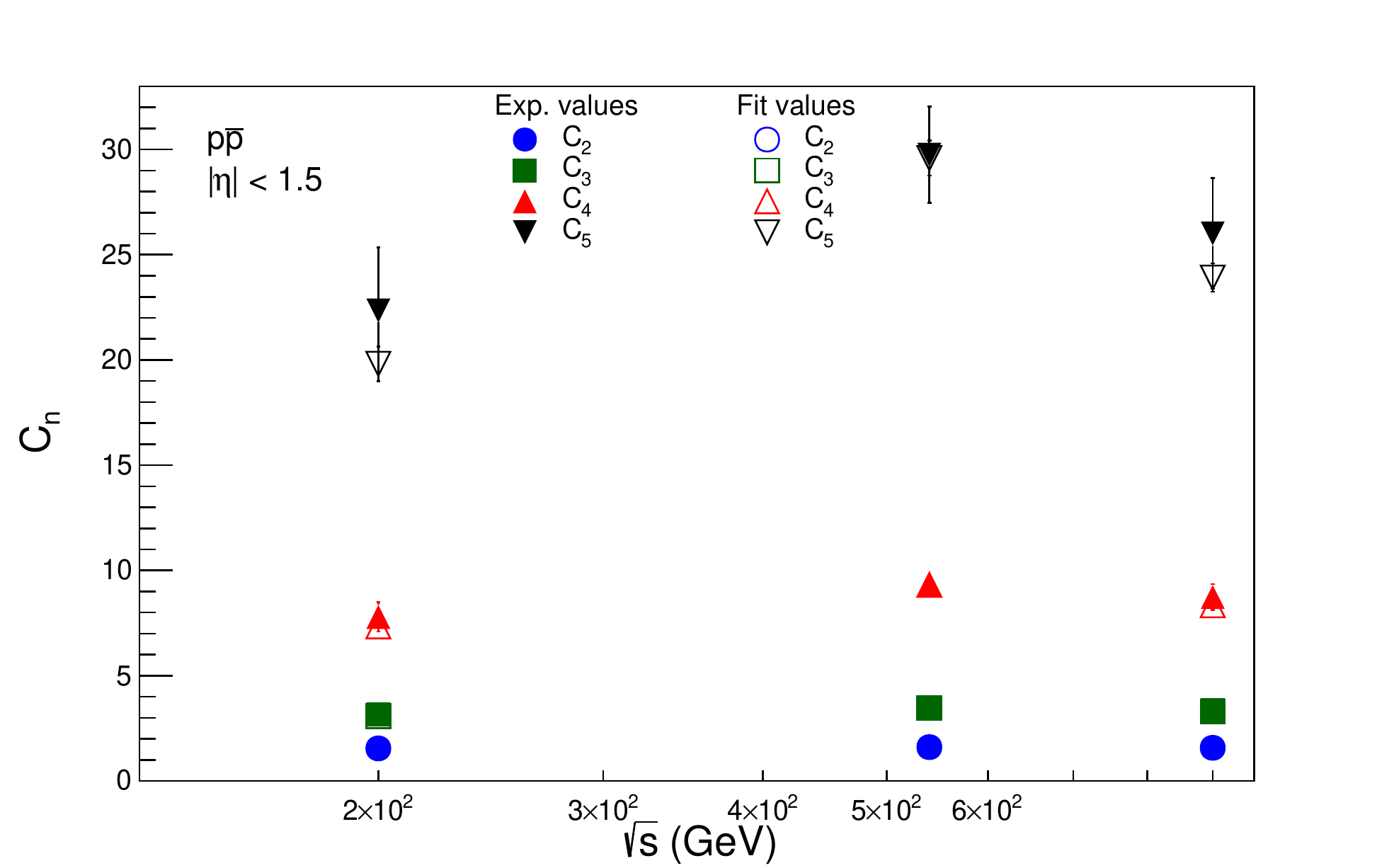}
\includegraphics[width=4.8 in, height =2.43 in]{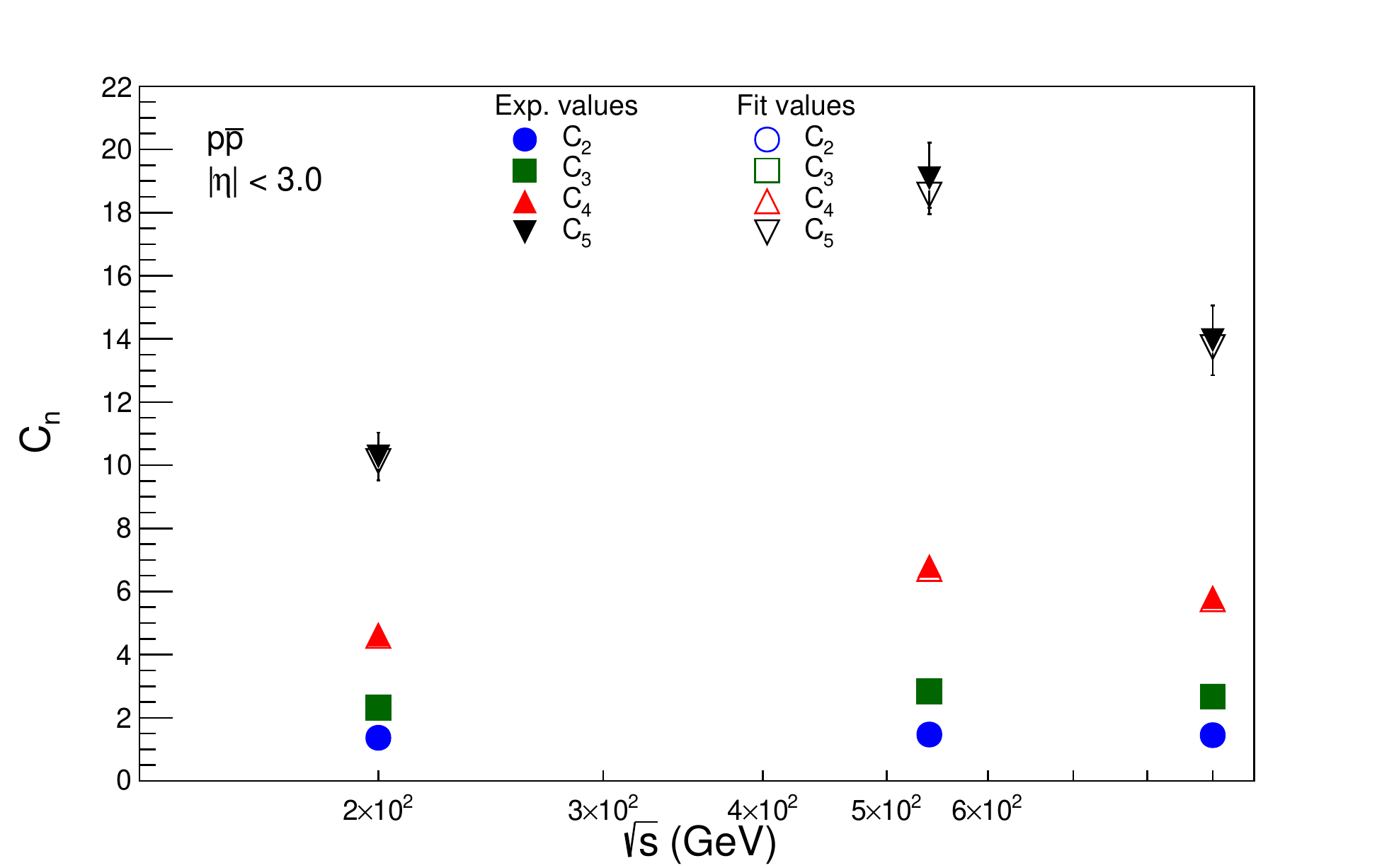}
\includegraphics[width=4.8 in, height =2.43 in]{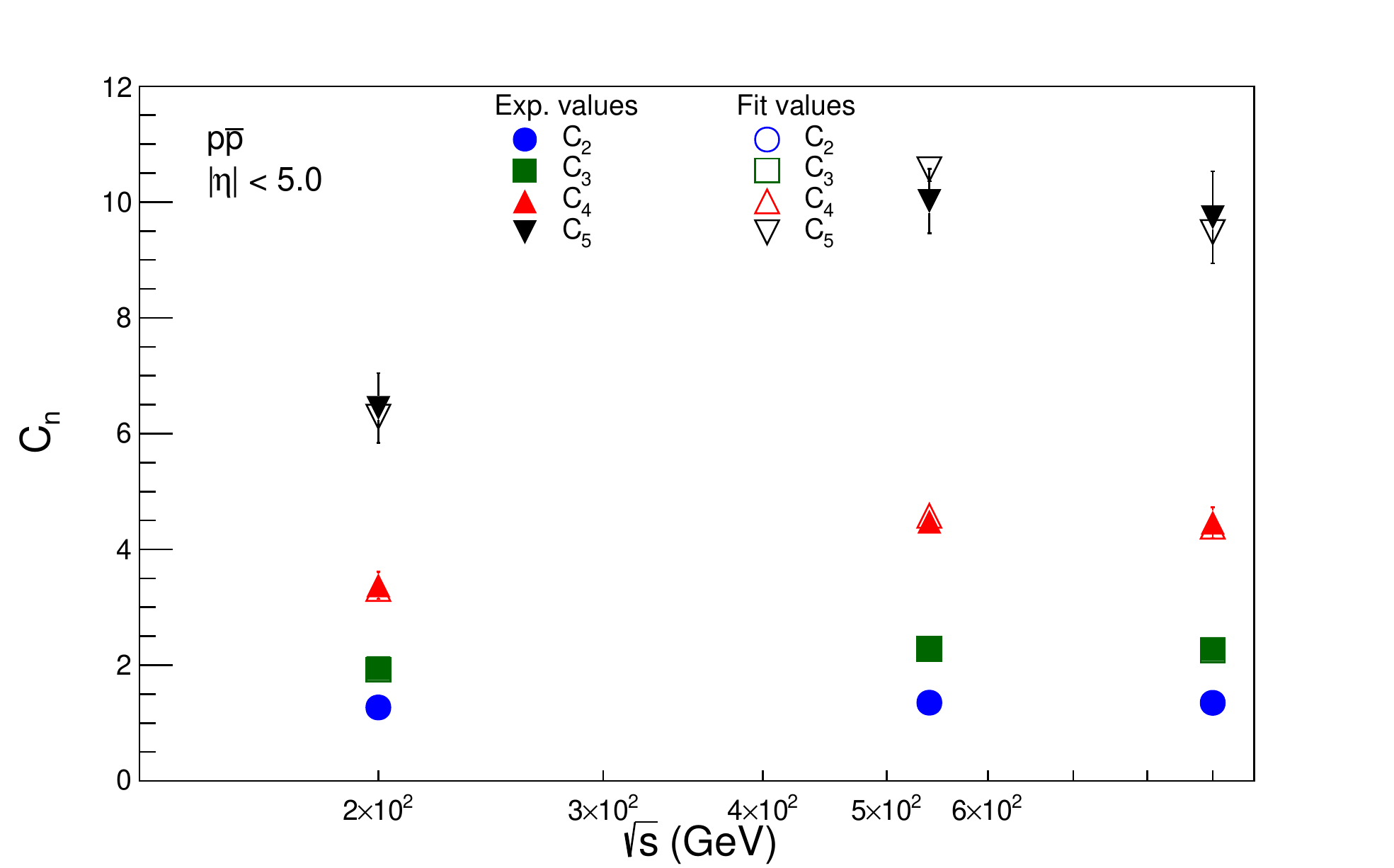}
\caption{Normalized factorial moments of modified shifted Gompertz distribution for the  $p\overline{p}$ collisions recorded by the $UA5$ experiment at different energies and in different rapidity bins.}
\end{figure}

\begin{figure}[ht]
\includegraphics[width=4.8 in, height =2.6 in]{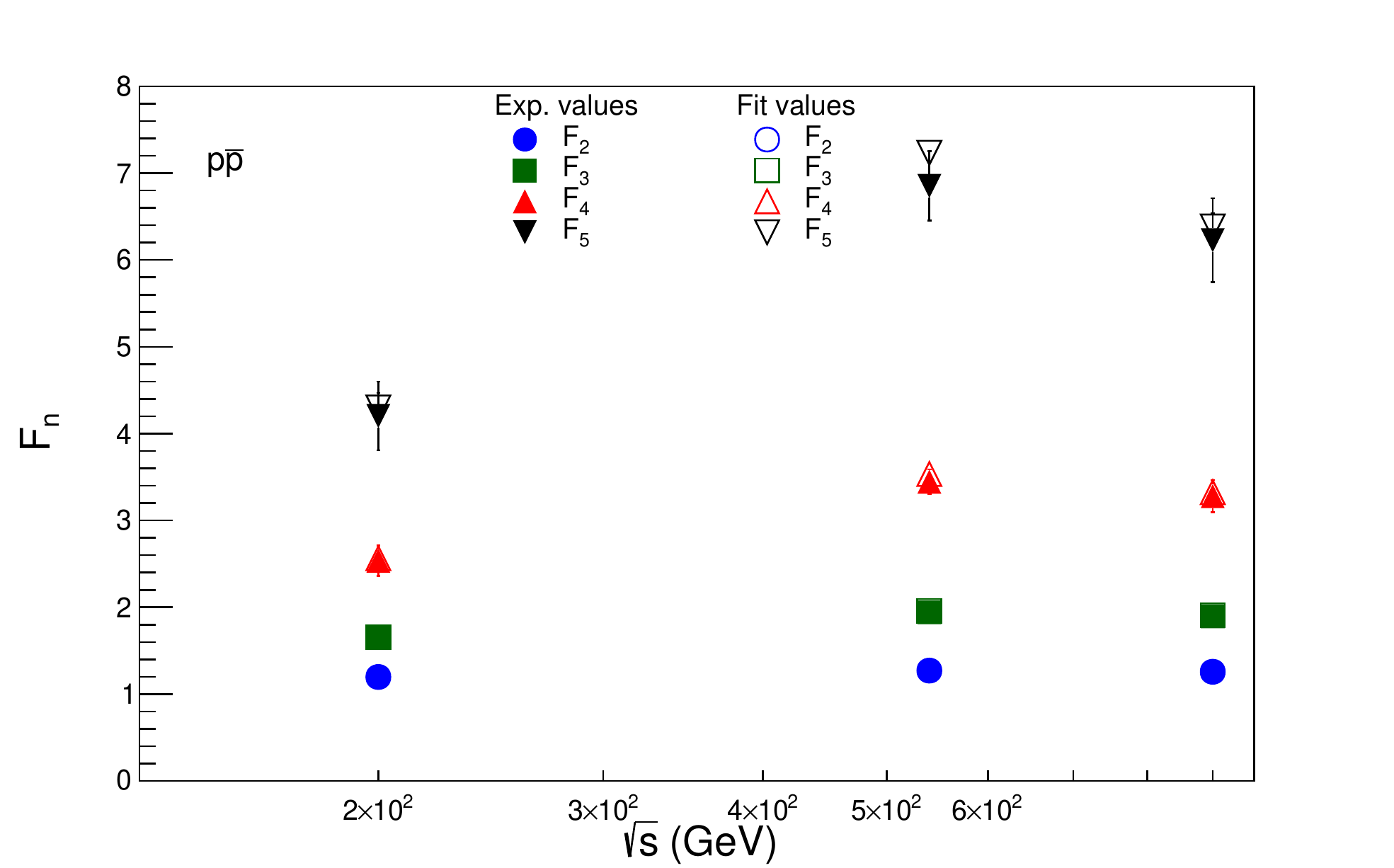}
\includegraphics[width=4.8 in, height =2.6 in]{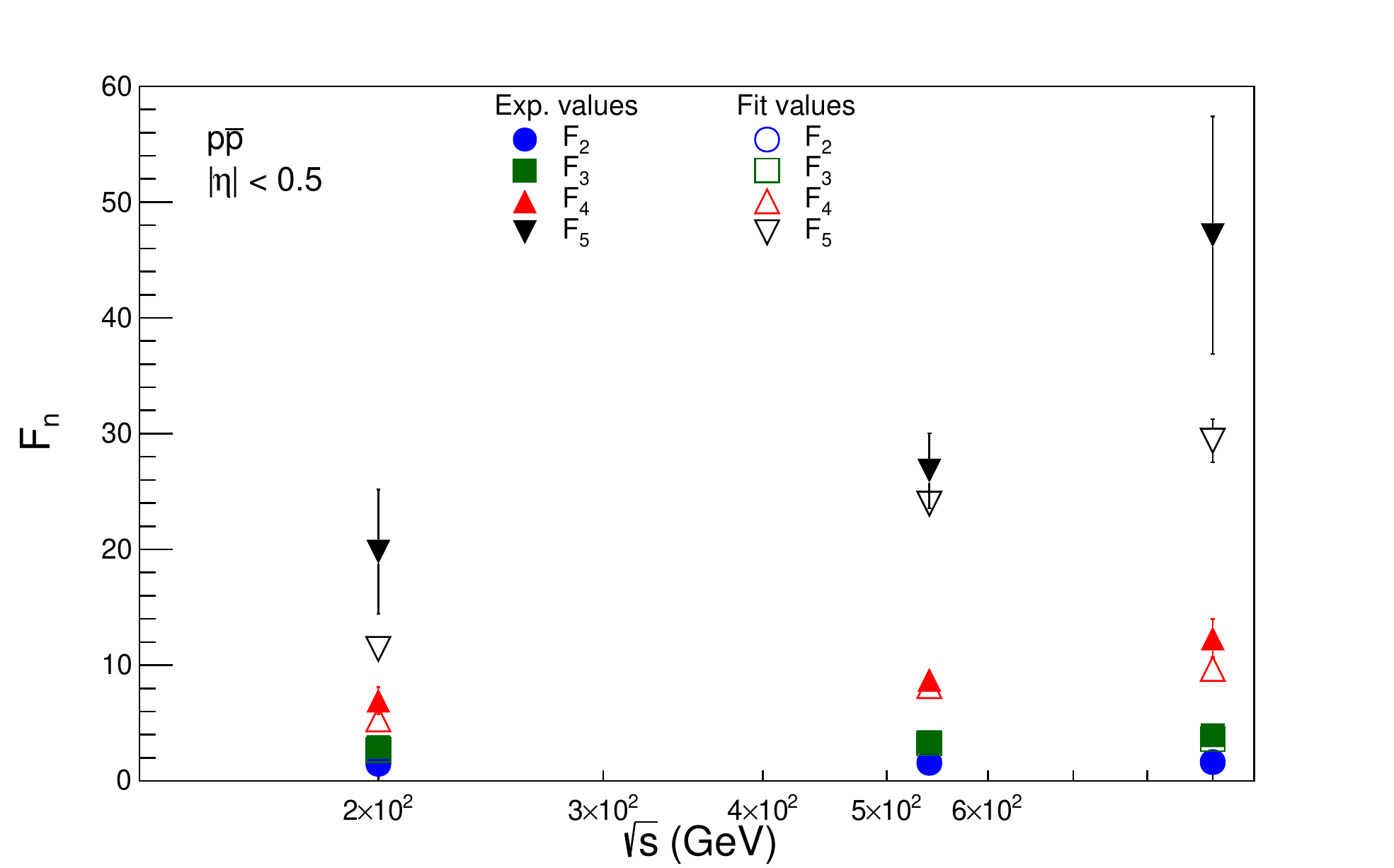}
\caption{Normalized factorial moments of modified shifted Gompertz distribution for the  $p\overline{p}$ collisions recorded by the  $UA5$ experiment at different energies and in different rapidity bins.}
\end{figure}
\begin{figure}[ht]
\includegraphics[width=4.8 in, height =2.43 in]{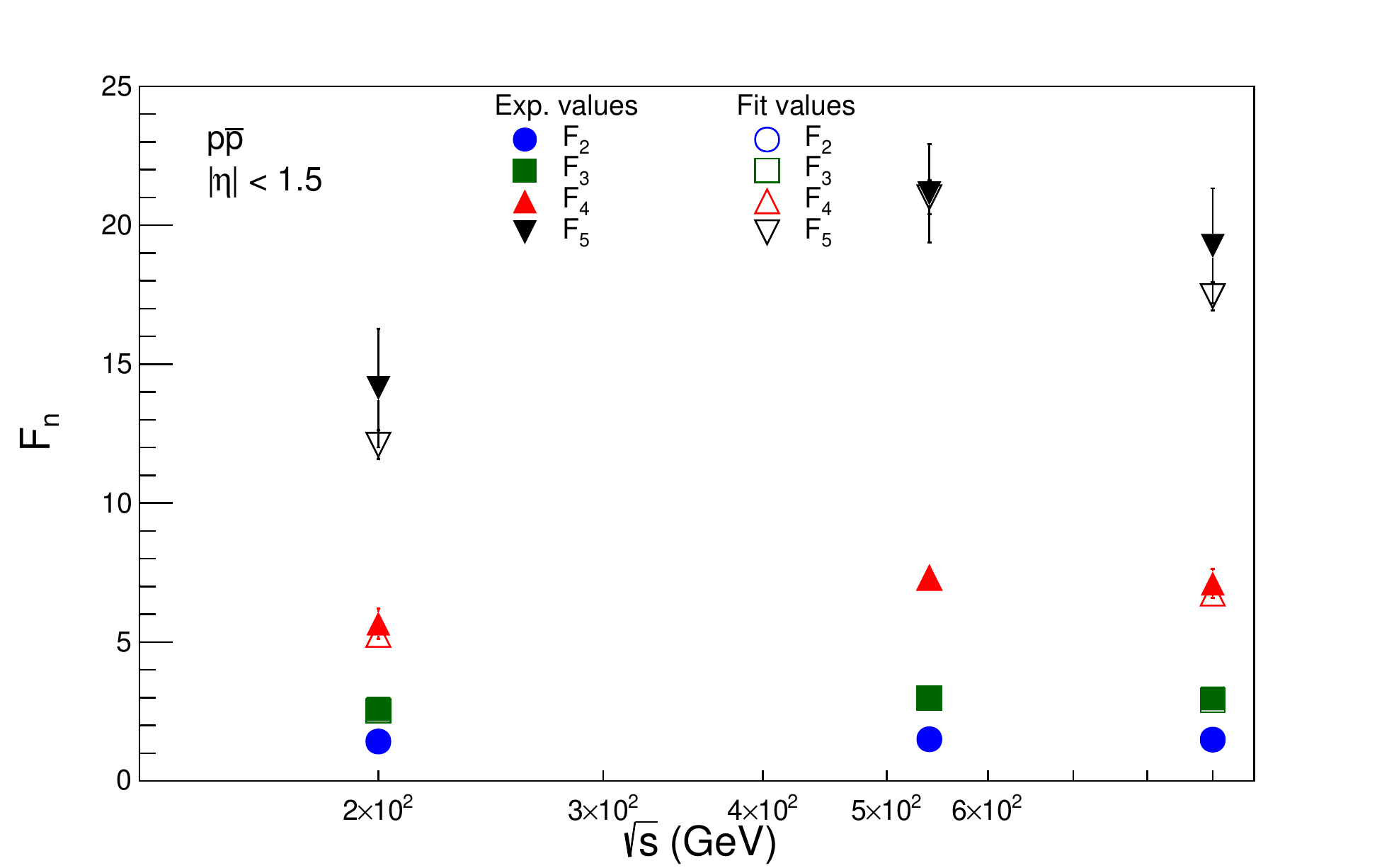}
\includegraphics[width=4.8 in, height =2.43 in]{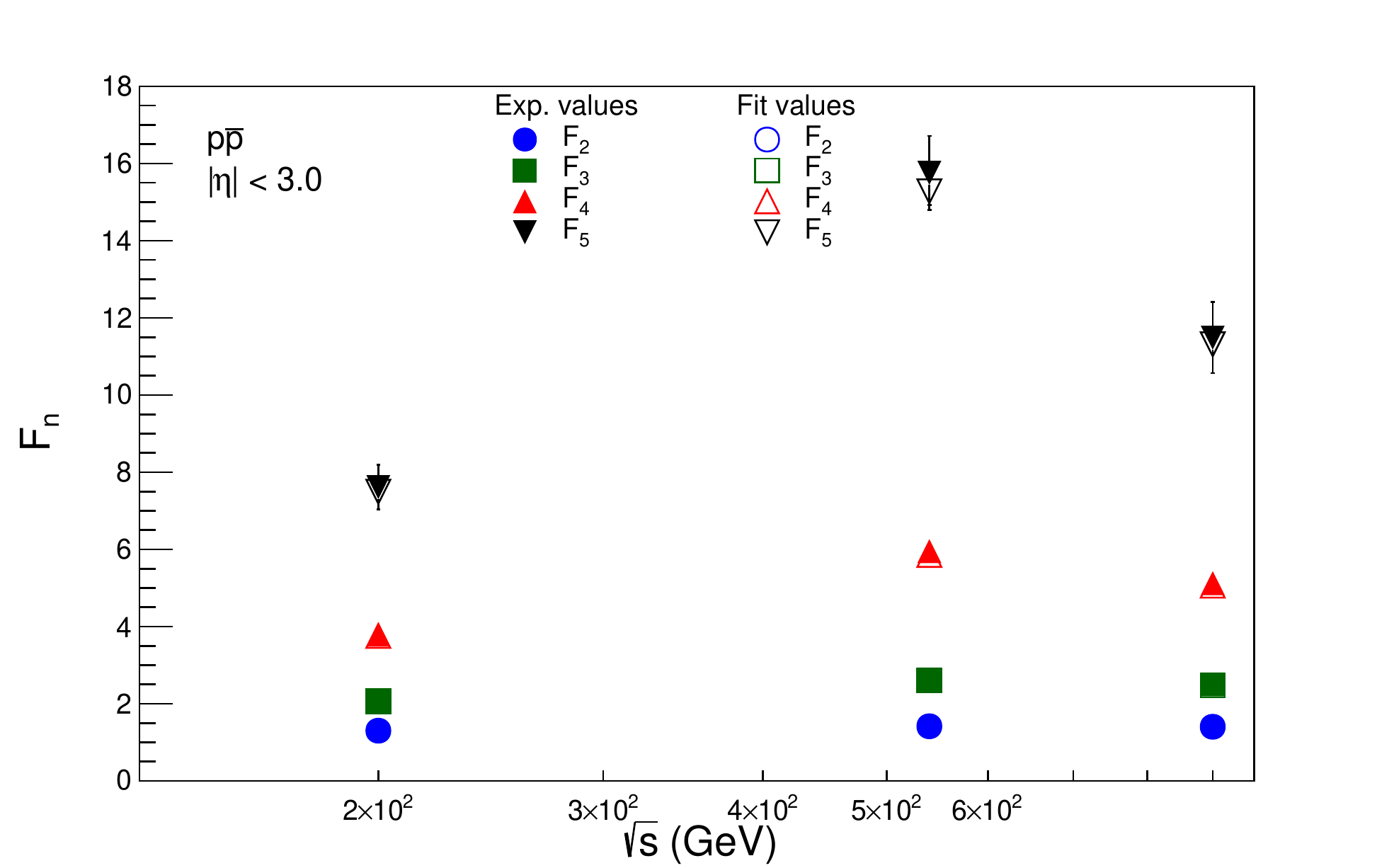}
\includegraphics[width=4.8 in, height =2.43 in]{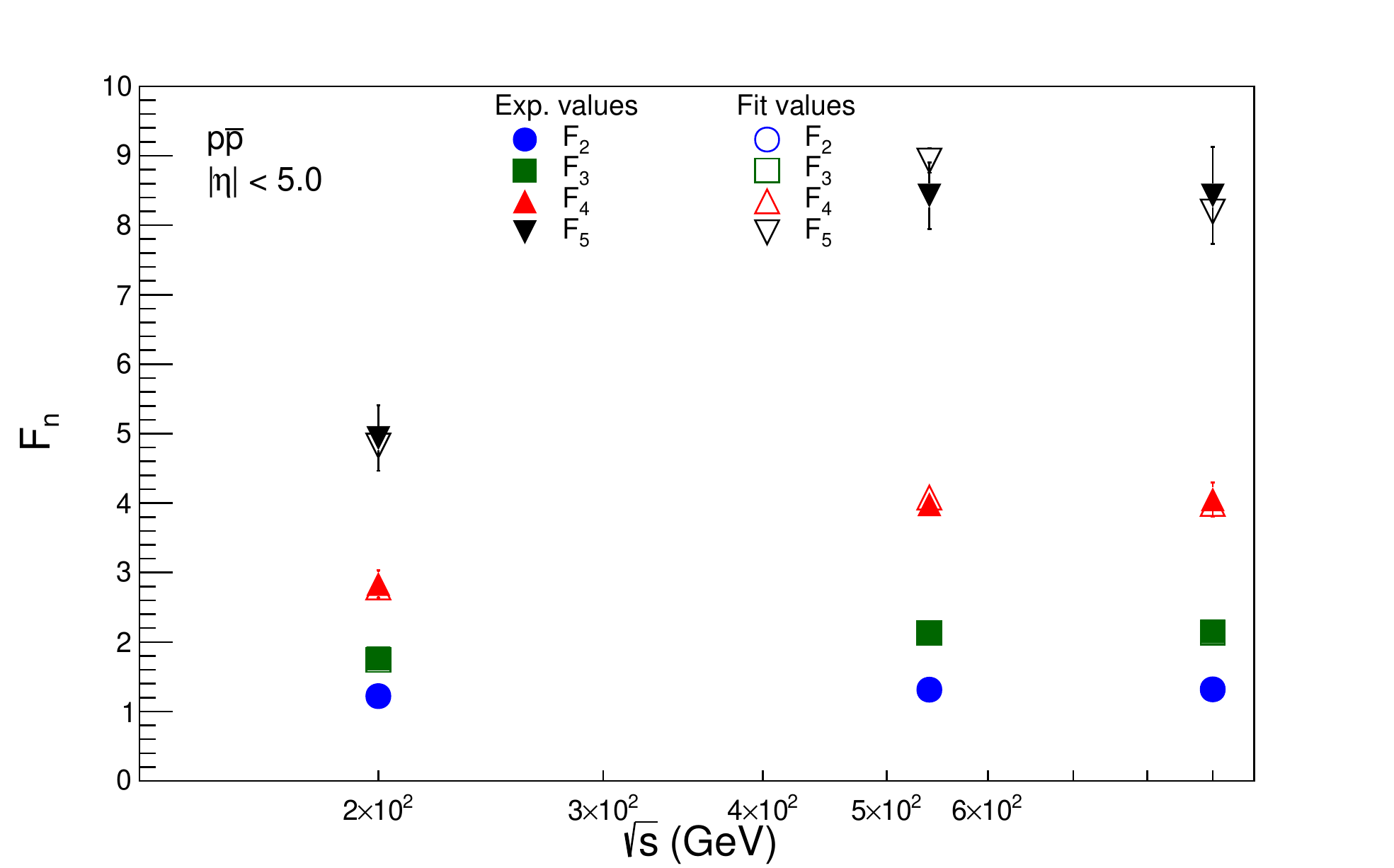}

\caption{Normalized factorial moments of modified shifted Gompertz distribution for the  $p\overline{p}$ collisions recorded by the $UA5$ experiment at different energies and in different rapidity bins.}
\end{figure}
\begin{figure}[ht]
\includegraphics[width=4.8 in, height =2.6 in]{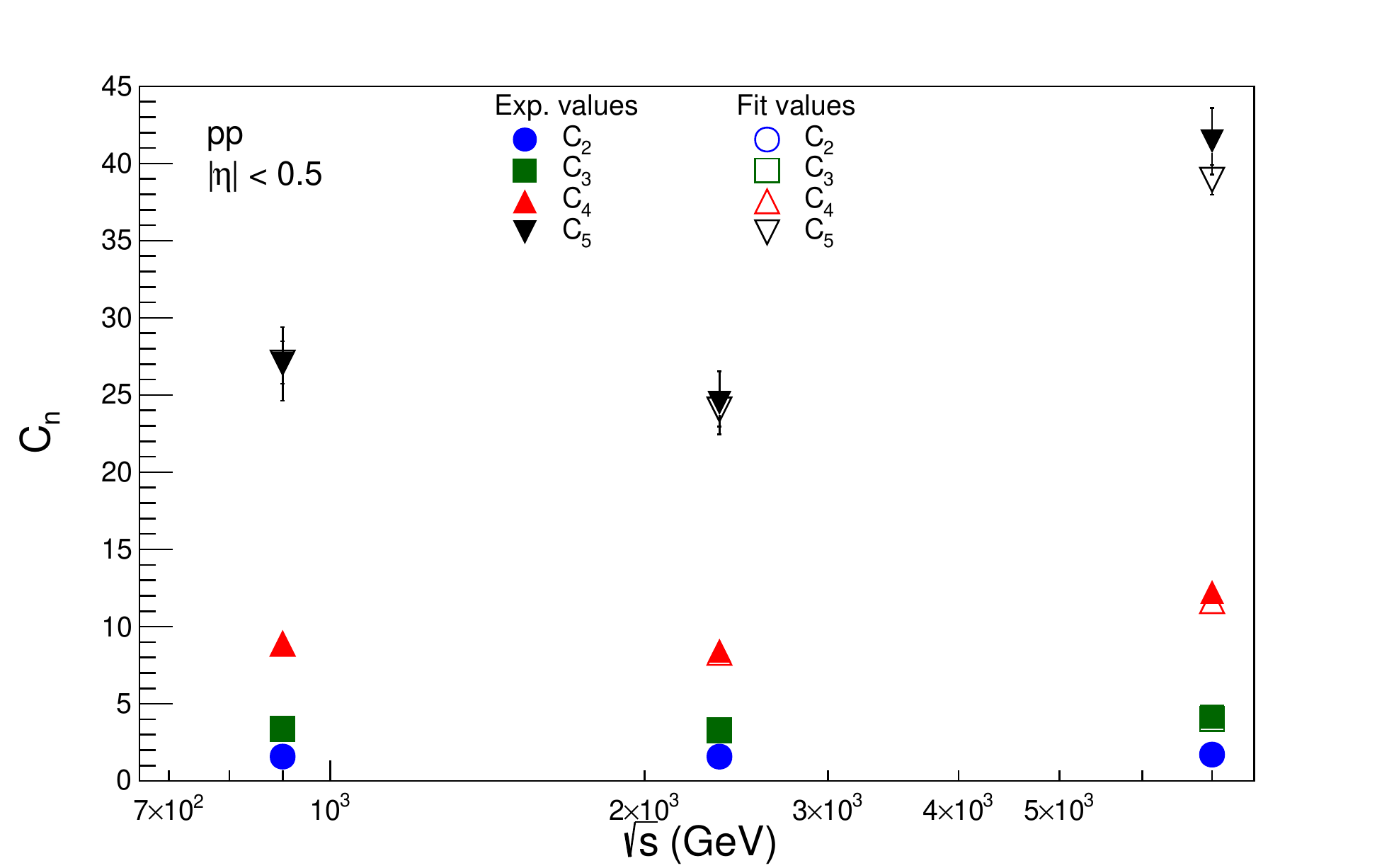}
\includegraphics[width=4.8 in, height =2.6 in]{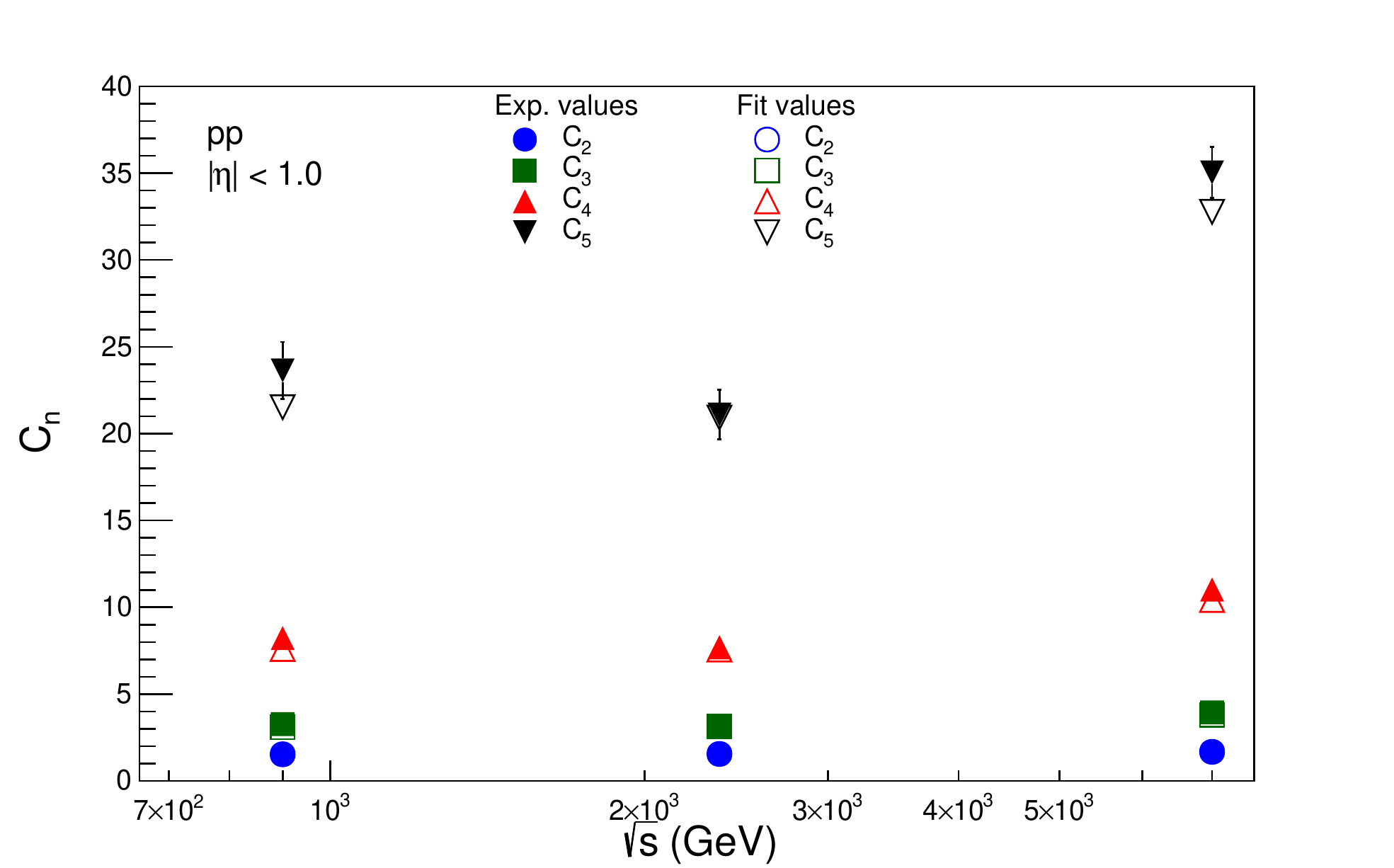}
\caption{Normalized moments of modified shifted Gompertz distribution for the $pp$ collisions recorded by the CMS experiment at different energies and in different rapidity bins.}
\end{figure}

\begin{figure}[ht]
\includegraphics[width=4.8 in, height =2.43 in]{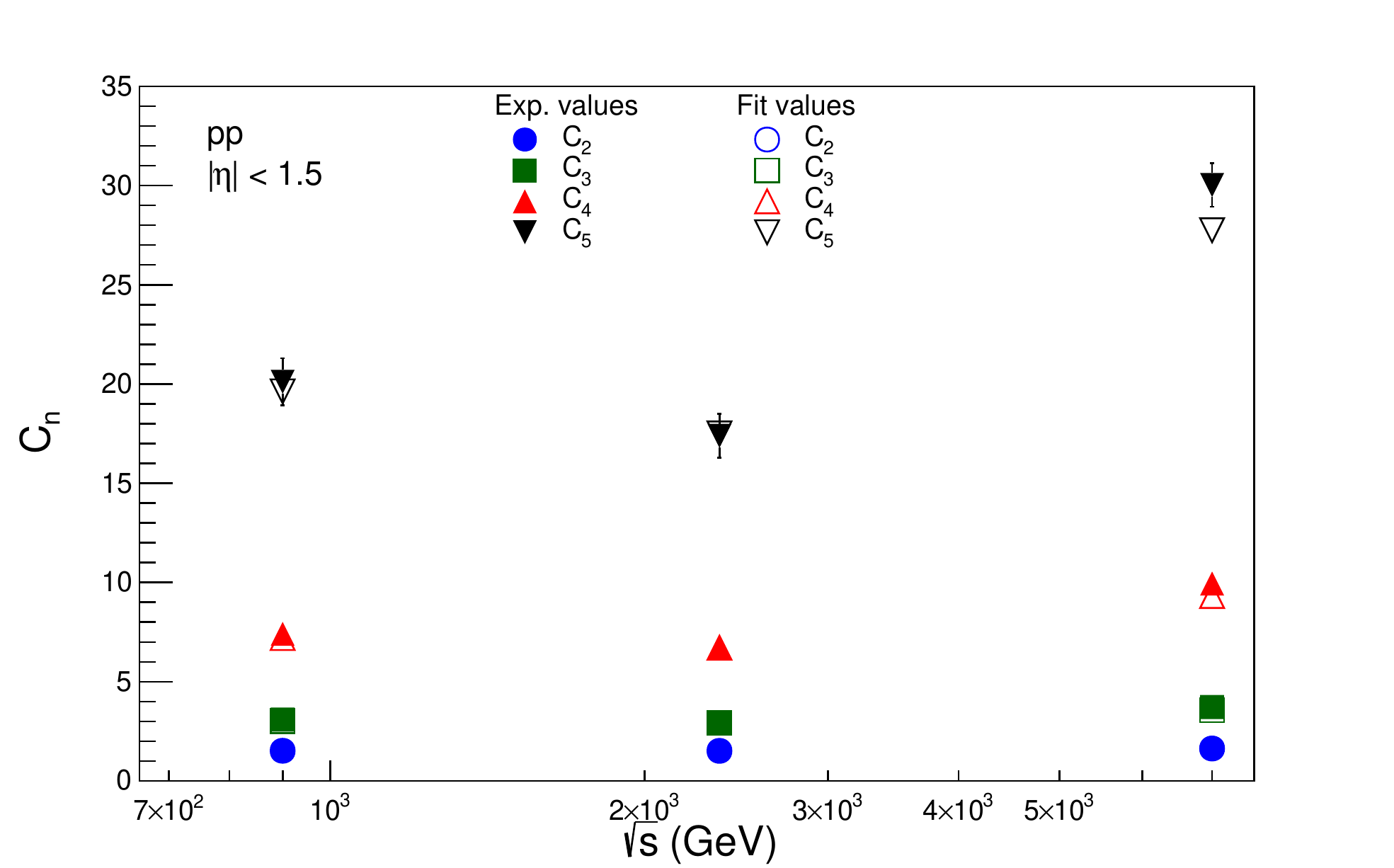}
\includegraphics[width=4.8 in, height =2.43 in]{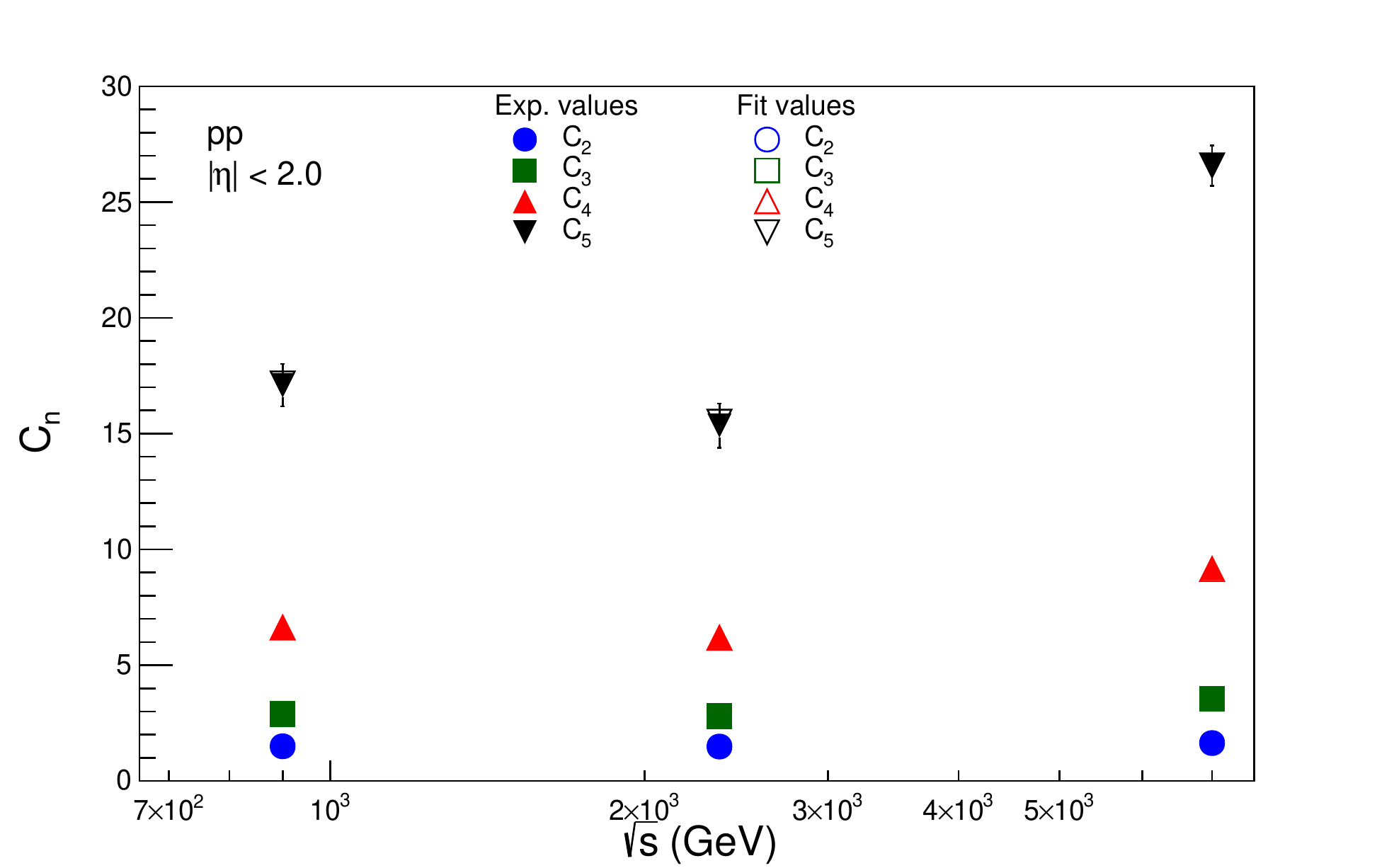}
\includegraphics[width=4.8 in, height =2.43 in]{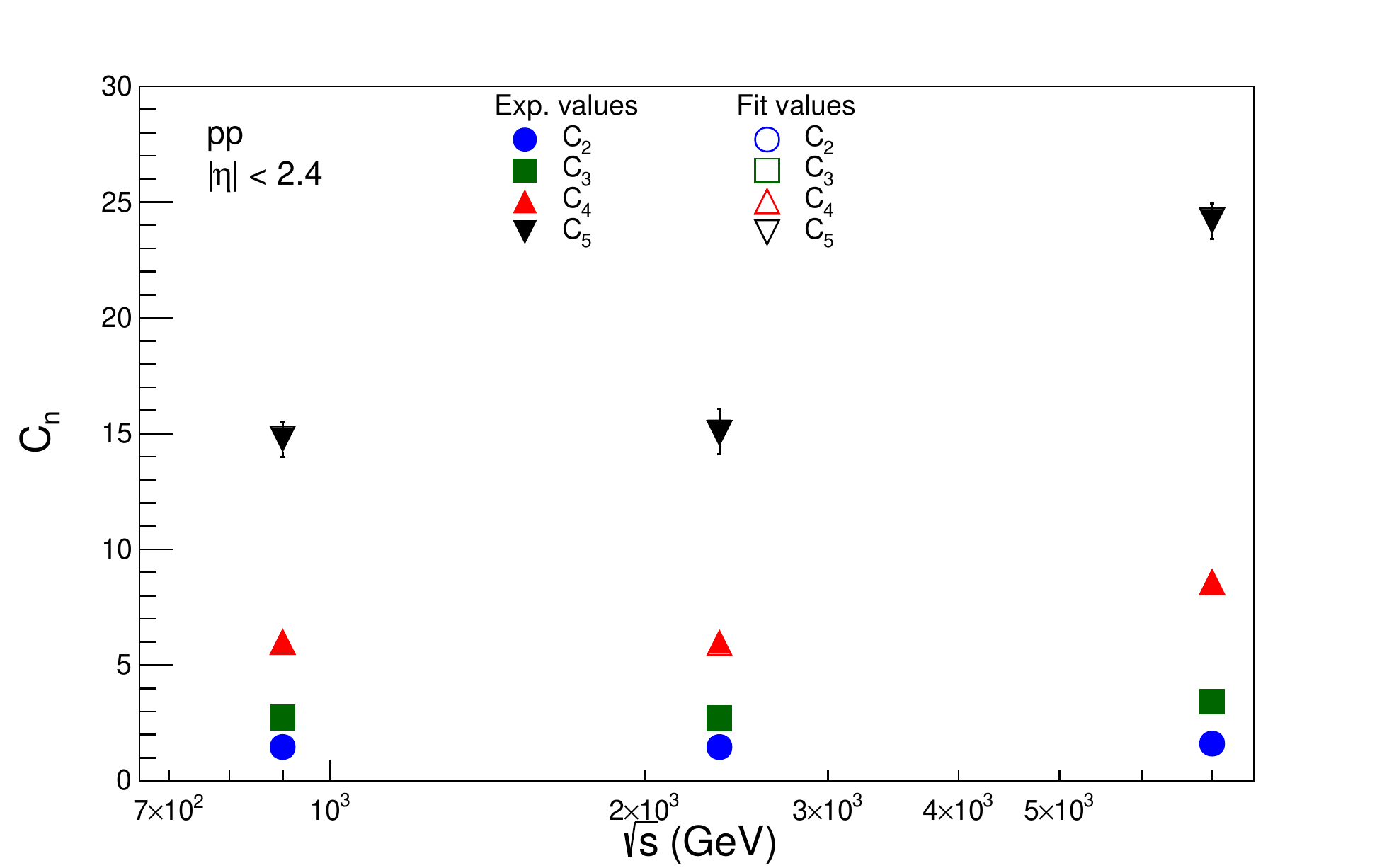}
\caption{Normalized moments of modified shifted Gompertz distribution for the $pp$ collisions recorded by the CMS experiment at different energies and in different rapidity bins.}
\end{figure}
\begin{figure}[ht]
\includegraphics[width=4.8 in, height =2.6 in]{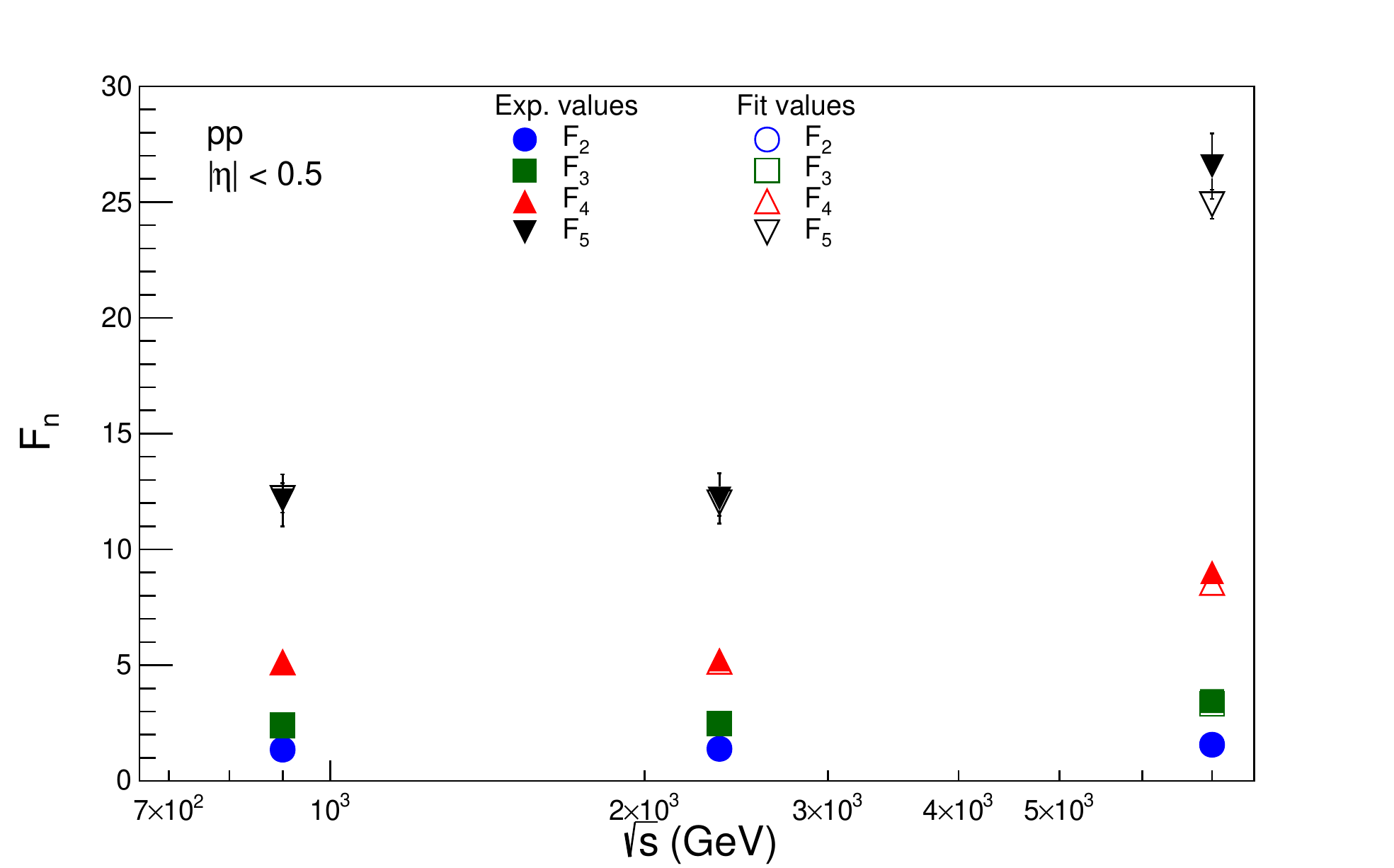}
\includegraphics[width=4.8 in, height =2.6 in]{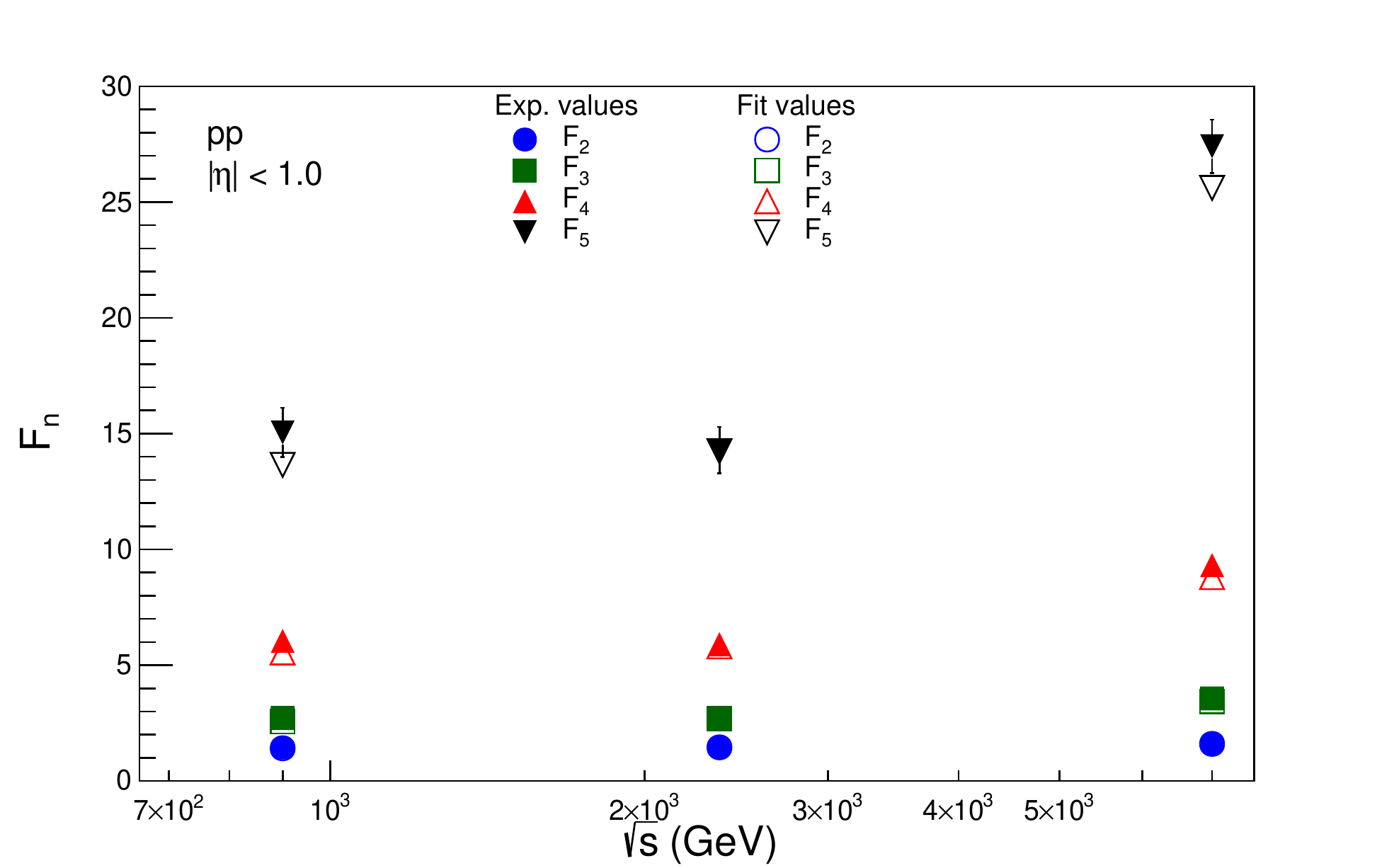}
\caption{Normalized factorial moments of modified shifted Gompertz distribution for the $pp$ collisions recorded by the CMS experiment at different energies and in different rapidity bins.}
\end{figure}
\begin{figure}[ht]
\includegraphics[width=4.8 in, height =2.43 in]{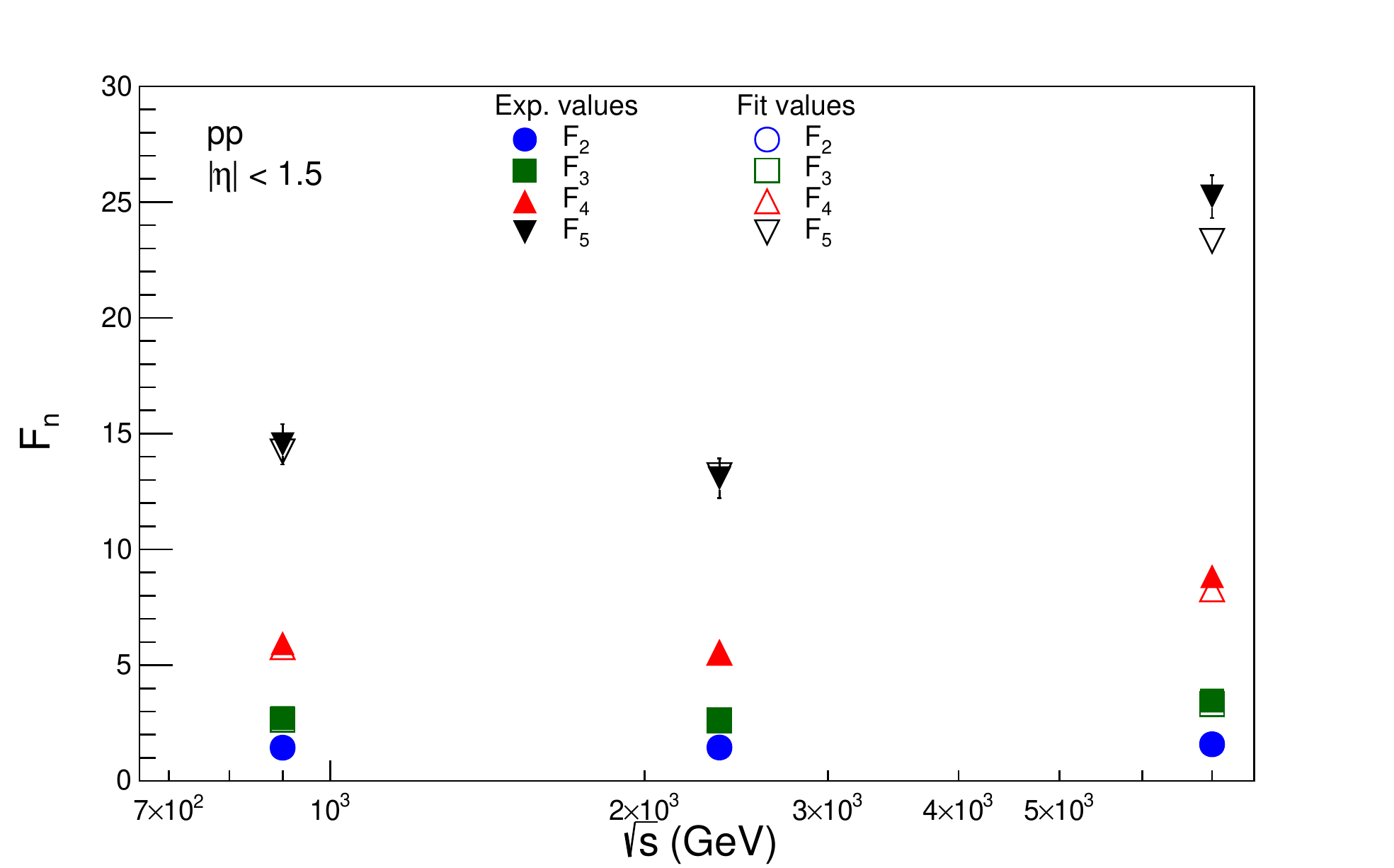}
\includegraphics[width=4.8 in, height =2.43 in]{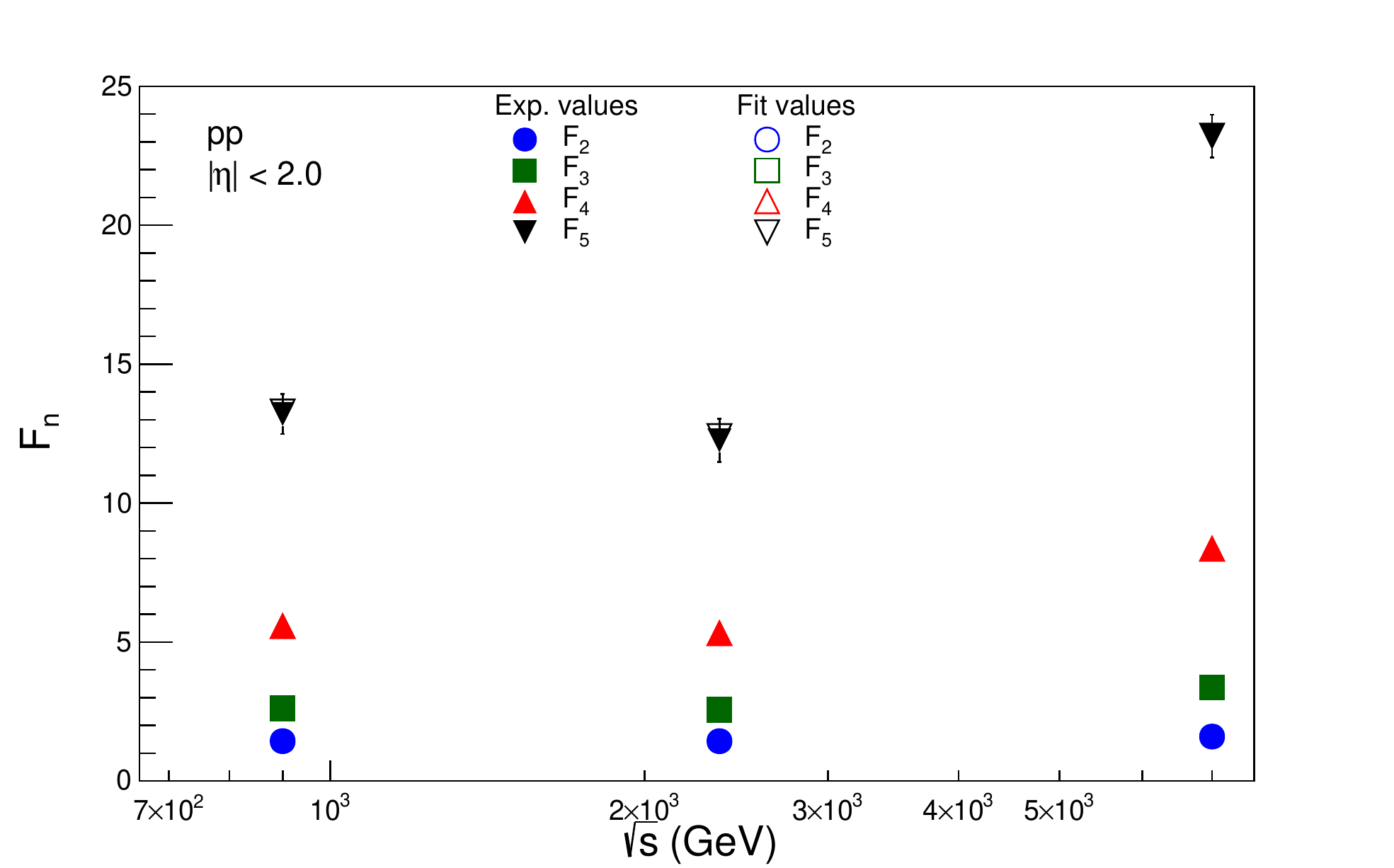}
\includegraphics[width=4.8 in, height =2.43 in]{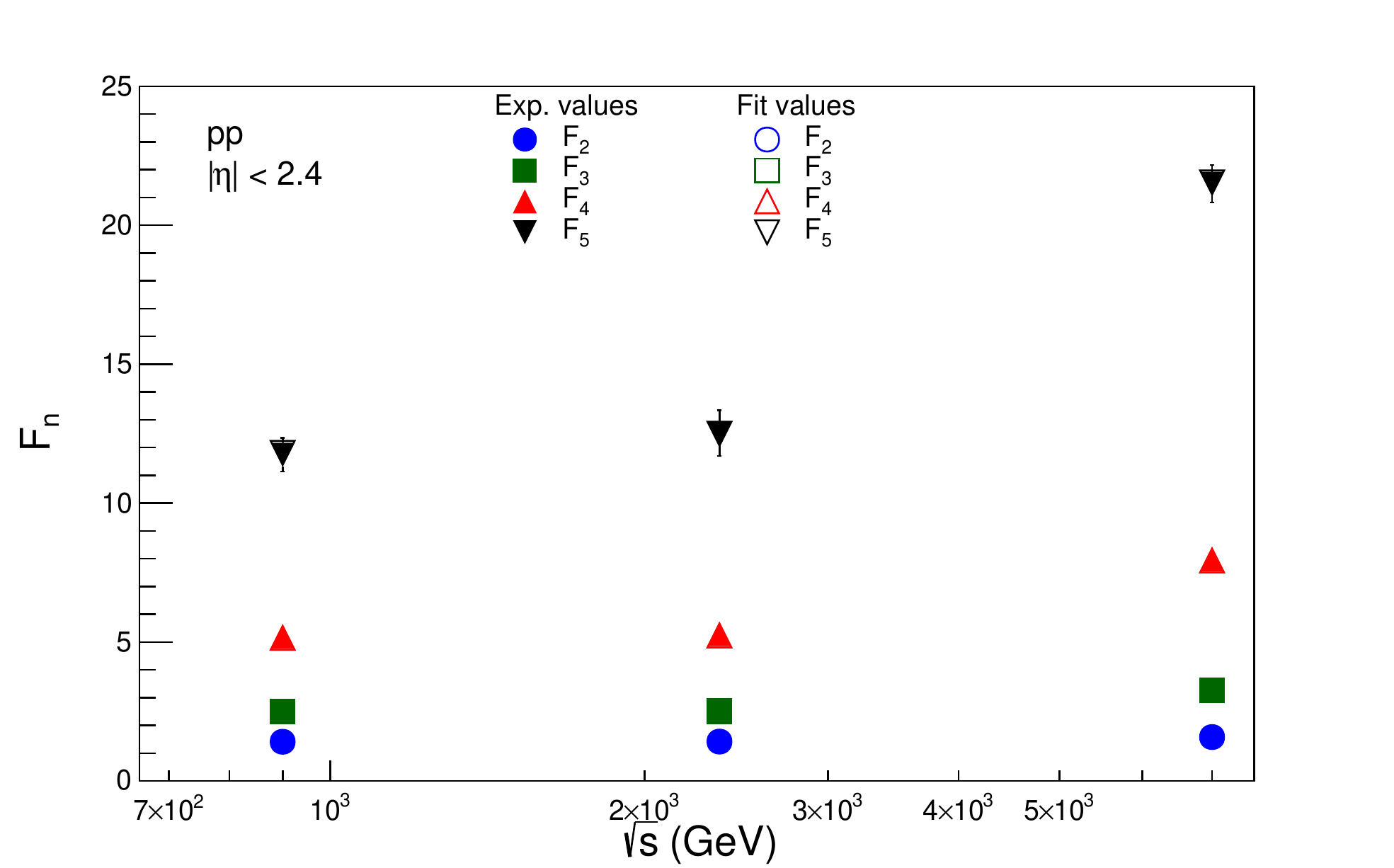}
\caption{Normalized factorial moments of modified shifted Gompertz distribution for the $pp$ collisions recorded by the CMS experiment at different energies and in different rapidity bins.}
\end{figure}

\begin{figure}[ht]
\includegraphics[width=4.8 in, height=2.6 in]{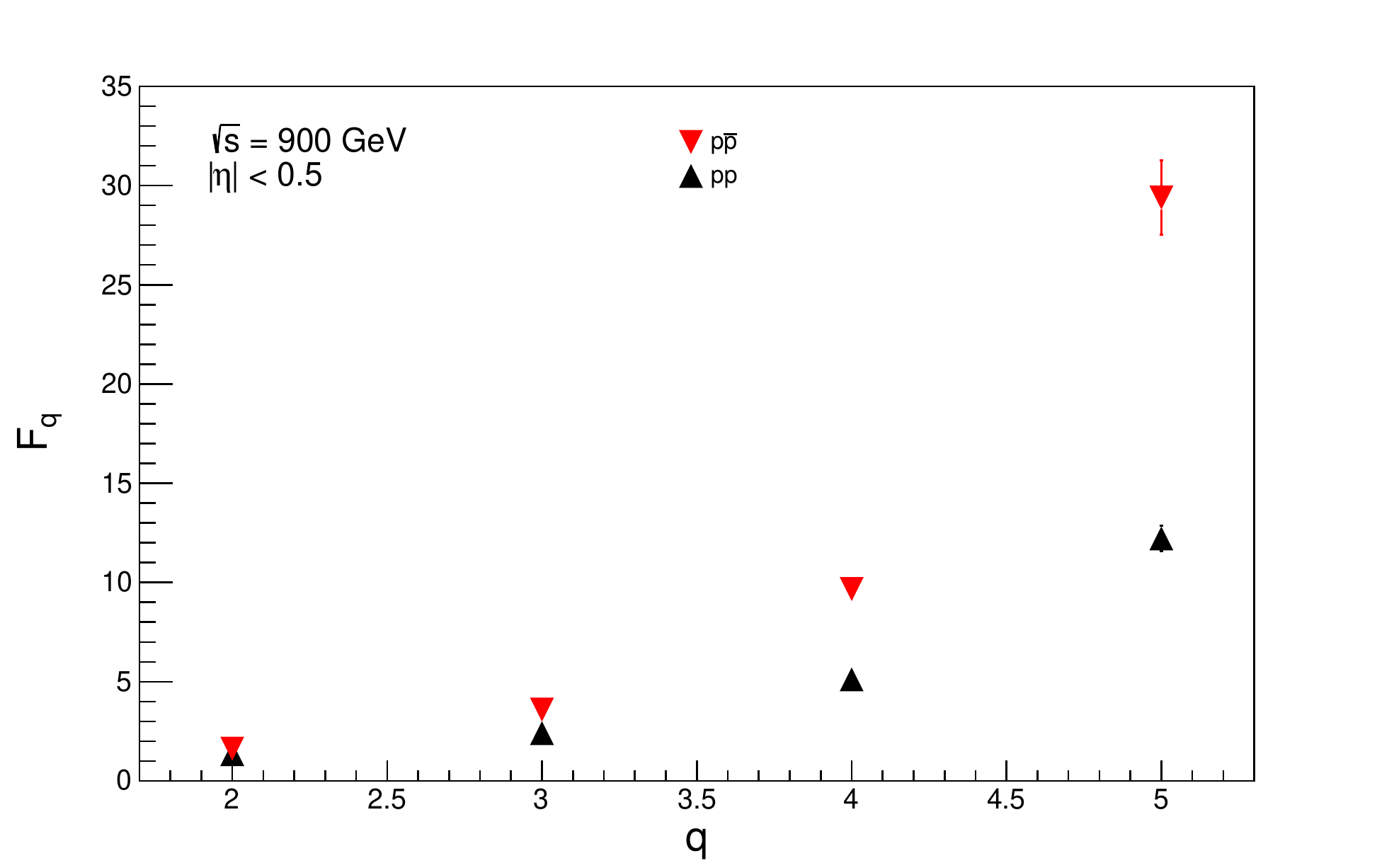}
\caption{Comparison of normalised factorial moments for $pp$ and $p\overline{p}$ interactions at the same c.m. energy in the same rapidity window.}
\end{figure}

\begin{figure}[ht]
\includegraphics[width=4.8 in, height =2.6 in]{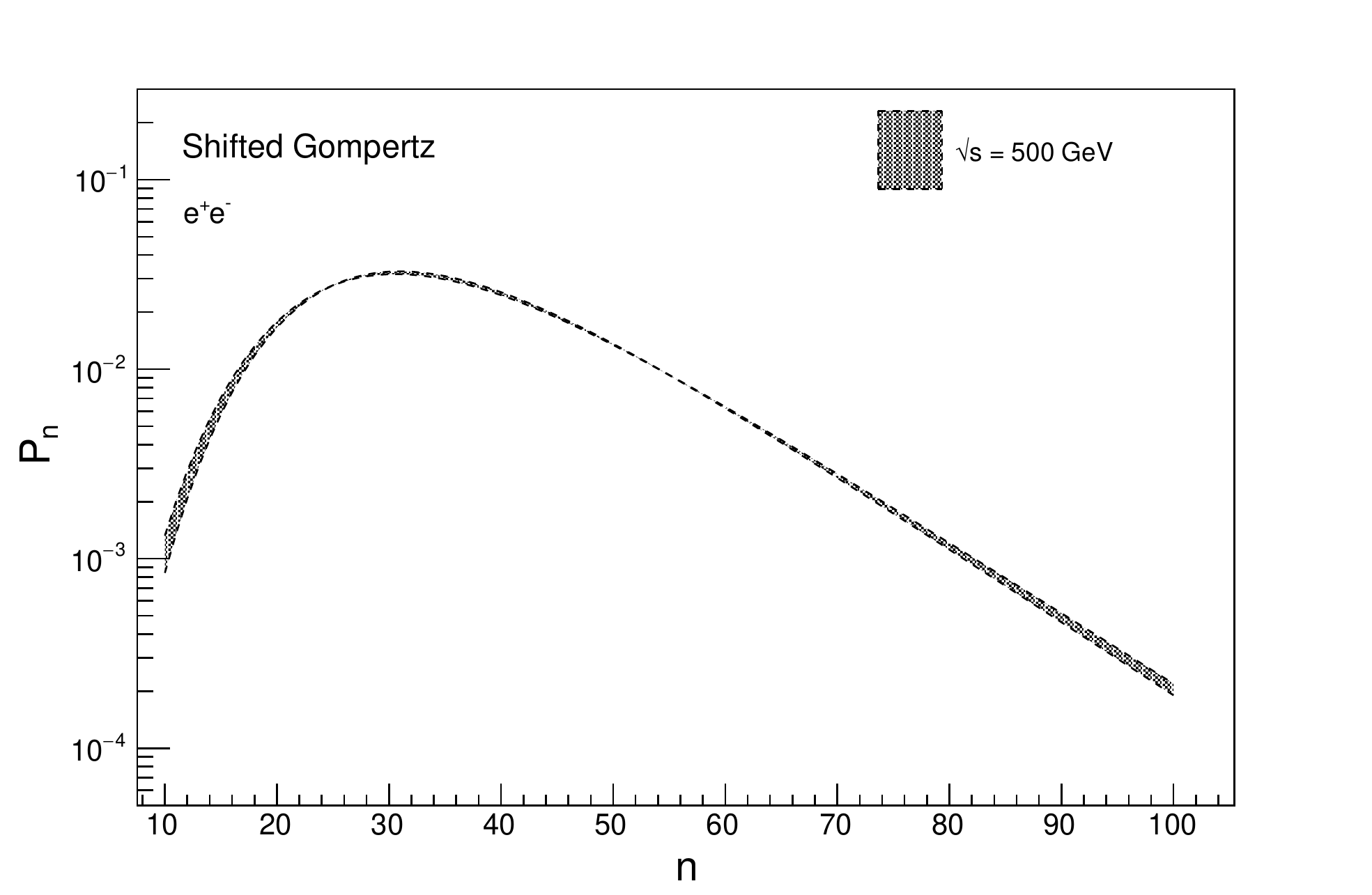}
\caption{Probability distribution predicted from shifted Gompertz distribution, in 1$\sigma$ confidence interval band, for the $e^+e^-$ collisions at 500 GeV.}
\end{figure}

\begin{table*}[t]
{\
\begin{adjustbox}{max width=\textwidth}

\begin{tabular}{|c|c|c|c|c|c|c|c|c|c|c|}
\hline
&  & &\multicolumn{4}{c|}{}&\multicolumn{4}{c|}{} \\

& Energy& &\multicolumn{4}{c|}{Normalized moments (Experiment)}&\multicolumn{4}{c|}{Normalized factorial moments (Experiment) } \\             

&(GeV)&$<n>$ &\multicolumn{4}{c|}{}&\multicolumn{4}{c|}{}  \\\cline{4-11}

&($e^{+}e^{-}$) & & & & & & & & &\\
& &  & C$_2$ & C$_3$ & C$_4$ & C$_5$ & F$_2$ & F$_3$ & F$_4$ & F$_5$                 \\\hline

&91 &20.464 $\pm$ 0.076 & 1.093 $\pm$ 0.004 & 1.296 $\pm$ 0.008 & 1.656 $\pm$ 0.016 & 2.268 $\pm$ 0.030 & 1.044 $\pm$ 0.003 & 1.141 $\pm$ 0.007 & 1.304 $\pm$ 0.013 & 1.560 $\pm$ 0.021  \\\cline{2-11}

&131 &23.301 $\pm$ 1.002 & 1.080 $\pm$ 0.044 & 1.252 $\pm$ 0.104 & 1.546 $\pm$ 0.195 & 2.016 $\pm$ 0.343 & 1.038 $\pm$ 0.043 & 1.117 $\pm$ 0.093 & 1.245 $\pm$ 0.157 & 1.430 $\pm$ 0.245  \\\cline{2-11}

&136 &24.012 $\pm$ 1.116 & 1.094 $\pm$ 0.051 & 1.295 $\pm$ 0.124 & 1.638 $\pm$ 0.237 & 2.194 $\pm$ 0.428 & 1.052 $\pm$ 0.050 & 1.162 $\pm$ 0.111 & 1.335 $\pm$ 0.195 & 1.586 $\pm$ 0.312  \\\cline{2-11}

&172 &27.031 $\pm$ 1.821 & 1.095 $\pm$ 0.059 & 1.299 $\pm$ 0.142 & 1.657 $\pm$ 0.275 & 2.245 $\pm$ 0.504 & 1.058 $\pm$ 0.057 & 1.181 $\pm$ 0.128 & 1.384 $\pm$ 0.228 & 1.692 $\pm$ 0.378  \\\cline{2-11}

L3&182 &26.795 $\pm$ 0.863 & 1.091 $\pm$ 0.030 & 1.287 $\pm$ 0.070 & 1.626 $\pm$ 0.135 & 2.180 $\pm$ 0.245 & 1.054 $\pm$ 0.028 & 1.168 $\pm$ 0.063 & 1.355 $\pm$ 0.112 & 1.633 $\pm$ 0.184  \\\cline{2-11}

&188 &26.653 $\pm$ 0.609 & 1.084 $\pm$ 0.020 & 1.260 $\pm$ 0.045 & 1.557 $\pm$ 0.085 & 2.026 $\pm$ 0.149 & 1.046 $\pm$ 0.019 & 1.141 $\pm$ 0.041 & 1.290 $\pm$ 0.070 & 1.502 $\pm$ 0.110  \\\cline{2-11}

&194 &27.178 $\pm$ 0.847 & 1.093 $\pm$ 0.027 & 1.295 $\pm$ 0.064 & 1.644 $\pm$ 0.123 & 2.216 $\pm$ 0.224 & 1.057 $\pm$ 0.026 & 1.177 $\pm$ 0.058 & 1.375 $\pm$ 0.102 & 1.670 $\pm$ 0.168  \\\cline{2-11}

&200 &27.575 $\pm$ 0.852 & 1.093 $\pm$ 0.027 & 1.294 $\pm$ 0.064 & 1.644 $\pm$ 0.124 & 2.215 $\pm$ 0.227 & 1.057 $\pm$ 0.026 & 1.178 $\pm$ 0.058 & 1.378 $\pm$ 0.103 & 1.676 $\pm$ 0.171  \\\cline{2-11}

& 206&27.917 $\pm$ 0.720 & 1.092 $\pm$ 0.021 & 1.290 $\pm$ 0.051 & 1.634 $\pm$ 0.098 & 2.195 $\pm$ 0.178 & 1.056 $\pm$ 0.021 & 1.176 $\pm$ 0.046 & 1.372 $\pm$ 0.082 & 1.665 $\pm$ 0.135  \\\hline

&91 &21.394 $\pm$ 0.219 & 1.092 $\pm$ 0.011 & 1.293 $\pm$ 0.027 & 1.648 $\pm$ 0.051 & 2.249 $\pm$ 0.095 & 1.045 $\pm$ 0.011 & 1.144 $\pm$ 0.024 & 1.311 $\pm$ 0.041 & 1.572 $\pm$ 0.067  \\\cline{2-11}

OPAL&161 &24.463 $\pm$ 0.859 & 1.095 $\pm$ 0.039 & 1.304 $\pm$ 0.093 & 1.673 $\pm$ 0.180 & 2.295 $\pm$ 0.333 & 1.054 $\pm$ 0.038 & 1.173 $\pm$ 0.084 & 1.373 $\pm$ 0.149 & 1.684 $\pm$ 0.247  \\\cline{2-11}

&183 &26.856 $\pm$ 0.754 & 1.097 $\pm$ 0.028 & 1.312 $\pm$ 0.068 & 1.696 $\pm$ 0.133 & 2.348 $\pm$ 0.252 & 1.060 $\pm$ 0.027 & 1.192 $\pm$ 0.062 & 1.419 $\pm$ 0.112 & 1.777 $\pm$ 0.195  \\\cline{2-11}

& 189&26.886 $\pm$ 0.505 & 1.096 $\pm$ 0.018 & 1.304 $\pm$ 0.043 & 1.669 $\pm$ 0.083 & 2.280 $\pm$ 0.153 & 1.059 $\pm$ 0.018 & 1.184 $\pm$ 0.039 & 1.395 $\pm$ 0.070 & 1.719 $\pm$ 0.117 \\\hline

\hline
&  & &\multicolumn{4}{c|}{}&\multicolumn{4}{c|}{} \\

& Energy& &\multicolumn{4}{c|}{Normalized moments (shifted Gompertz)}&\multicolumn{4}{c|}{Normalized factorial moments (shifted Gompertz) } \\             

&(GeV)&$<n>$ &\multicolumn{4}{c|}{}&\multicolumn{4}{c|}{}  \\\cline{4-11}

&($e^{+}e^{-}$) & & & & & & & & &\\
& &  & C$_2$ & C$_3$ & C$_4$ & C$_5$ & F$_2$ & F$_3$ & F$_4$ & F$_5$                 \\\hline

&91 & 20.467 $\pm$ 0.029 & 1.084 $\pm$ 0.002 & 1.278 $\pm$ 0.004 & 1.643 $\pm$ 0.008 & 2.307 $\pm$ 0.015 & 1.035 $\pm$ 0.002 & 1.124 $\pm$ 0.004 & 1.296 $\pm$ 0.006 & 1.604 $\pm$ 0.010  \\\cline{2-11}

& 131& 23.101 $\pm$ 0.325 & 1.084 $\pm$ 0.017 & 1.268 $\pm$ 0.039 & 1.591 $\pm$ 0.074 & 2.126 $\pm$ 0.131 & 1.041 $\pm$ 0.016 & 1.131 $\pm$ 0.035 & 1.284 $\pm$ 0.060 & 1.515 $\pm$ 0.094  \\\cline{2-11}

& 136& 23.530 $\pm$ 0.348 & 1.096 $\pm$ 0.018 & 1.308 $\pm$ 0.043 & 1.684 $\pm$ 0.083 & 2.317 $\pm$ 0.152 & 1.054 $\pm$ 0.017 & 1.172 $\pm$ 0.039 & 1.372 $\pm$ 0.068 & 1.680 $\pm$ 0.111  \\\cline{2-11}

& 172 & 26.706 $\pm$ 0.359 & 1.097 $\pm$ 0.016 & 1.309 $\pm$ 0.038 & 1.686 $\pm$ 0.073 & 2.317 $\pm$ 0.134 & 1.059 $\pm$ 0.015 & 1.189 $\pm$ 0.035 & 1.408 $\pm$ 0.062 & 1.747 $\pm$ 0.102  \\\cline{2-11}

L3&182 & 26.779 $\pm$ 0.259 & 1.096 $\pm$ 0.010 & 1.307 $\pm$ 0.025 & 1.679 $\pm$ 0.048 & 2.303 $\pm$ 0.089 & 1.059 $\pm$ 0.010 & 1.187 $\pm$ 0.023 & 1.403 $\pm$ 0.040 & 1.737 $\pm$ 0.067  \\\cline{2-11}

& 188& 26.624 $\pm$ 0.172 & 1.088 $\pm$ 0.007 & 1.276 $\pm$ 0.016 & 1.599 $\pm$ 0.030 & 2.121 $\pm$ 0.053 & 1.050 $\pm$ 0.006 & 1.156 $\pm$ 0.014 & 1.328 $\pm$ 0.025 & 1.581 $\pm$ 0.040  \\\cline{2-11}

& 194& 27.007 $\pm$ 0.232 & 1.095 $\pm$ 0.009 & 1.305 $\pm$ 0.022 & 1.674 $\pm$ 0.043 & 2.288 $\pm$ 0.078 & 1.058 $\pm$ 0.009 & 1.186 $\pm$ 0.020 & 1.400 $\pm$ 0.036 & 1.729 $\pm$ 0.060  \\\cline{2-11}

&200 & 27.599 $\pm$ 0.251 & 1.095 $\pm$ 0.009 & 1.300 $\pm$ 0.022 & 1.660 $\pm$ 0.043 & 2.253 $\pm$ 0.078 & 1.058 $\pm$ 0.009 & 1.184 $\pm$ 0.020 & 1.392 $\pm$ 0.036 & 1.709 $\pm$ 0.060  \\\cline{2-11}

&206 & 27.940 $\pm$ 0.209 & 1.093 $\pm$ 0.007 & 1.295 $\pm$ 0.018 & 1.646 $\pm$ 0.034 & 2.222 $\pm$ 0.062 & 1.057 $\pm$ 0.007 & 1.180 $\pm$ 0.016 & 1.383 $\pm$ 0.029 & 1.688 $\pm$ 0.047  \\\hline

&91 & 21.547 $\pm$ 0.086 & 1.093 $\pm$ 0.005 & 1.306 $\pm$ 0.012 & 1.706 $\pm$ 0.023 & 2.425 $\pm$ 0.043 & 1.046 $\pm$ 0.005 & 1.159 $\pm$ 0.010 & 1.368 $\pm$ 0.018 & 1.726 $\pm$ 0.031  \\\cline{2-11}

OPAL&161 & 24.322 $\pm$ 0.325 & 1.099 $\pm$ 0.017 & 1.325 $\pm$ 0.040 & 1.742 $\pm$ 0.078 & 2.479 $\pm$ 0.148 & 1.058 $\pm$ 0.016 & 1.193 $\pm$ 0.036 & 1.436 $\pm$ 0.065 & 1.838 $\pm$ 0.110  \\\cline{2-11}

&183 & 26.980 $\pm$ 0.244 & 1.102 $\pm$ 0.011 & 1.333 $\pm$ 0.026 & 1.761 $\pm$ 0.051 & 2.517 $\pm$ 0.097 & 1.065 $\pm$ 0.010 & 1.213 $\pm$ 0.023 & 1.481 $\pm$ 0.043 & 1.926 $\pm$ 0.075  \\\cline{2-11}

&189 & 27.176 $\pm$ 0.164 & 1.103 $\pm$ 0.007 & 1.334 $\pm$ 0.017 & 1.750 $\pm$ 0.033 & 2.466 $\pm$ 0.062 & 1.066 $\pm$ 0.007 & 1.214 $\pm$ 0.016 & 1.471 $\pm$ 0.028 & 1.883 $\pm$ 0.048  \\\cline{2-11}

\hline
&  & &\multicolumn{4}{c|}{}&\multicolumn{4}{c|}{} \\

& Energy& &\multicolumn{4}{c|}{Normalized moments (Modified shifted Gompertz)}&\multicolumn{4}{c|}{Normalized factorial moments (Modified shifted Gompertz) } \\             

&(GeV)&$<n>$ &\multicolumn{4}{c|}{}&\multicolumn{4}{c|}{}  \\\cline{4-11}

&($e^{+}e^{-}$) & & & & & & & & &\\
&  &  & C$_2$ & C$_3$ & C$_4$ & C$_5$ & F$_2$ & F$_3$ & F$_4$ & F$_5$                 \\\hline

& 91& 20.209 $\pm$ 0.031 & 1.092 $\pm$ 0.002 & 1.297 $\pm$ 0.005 & 1.670 $\pm$ 0.009 & 2.329 $\pm$ 0.016 & 1.043 $\pm$ 0.002 & 1.139 $\pm$ 0.004 & 1.313 $\pm$ 0.007 & 1.607 $\pm$ 0.011  \\\cline{2-11}

& 131& 23.293 $\pm$ 0.427 & 1.081 $\pm$ 0.022 & 1.255 $\pm$ 0.050 & 1.555 $\pm$ 0.093 & 2.042 $\pm$ 0.164 & 1.038 $\pm$ 0.021 & 1.120 $\pm$ 0.045 & 1.253 $\pm$ 0.076 & 1.451 $\pm$ 0.117  \\\cline{2-11}

&136 & 23.867 $\pm$ 0.499 & 1.092 $\pm$ 0.025 & 1.291 $\pm$ 0.059 & 1.635 $\pm$ 0.113 & 2.198 $\pm$ 0.203 & 1.050 $\pm$ 0.024 & 1.157 $\pm$ 0.054 & 1.331 $\pm$ 0.093 & 1.589 $\pm$ 0.148  \\\cline{2-11}

& 172& 26.852 $\pm$ 0.458 & 1.092 $\pm$ 0.020 & 1.293 $\pm$ 0.048 & 1.645 $\pm$ 0.091 & 2.229 $\pm$ 0.165 & 1.055 $\pm$ 0.019 & 1.174 $\pm$ 0.044 & 1.372 $\pm$ 0.077 & 1.677 $\pm$ 0.125  \\\cline{2-11}

L3& 182& 26.771 $\pm$ 0.317 & 1.093 $\pm$ 0.013 & 1.295 $\pm$ 0.030 & 1.650 $\pm$ 0.058 & 2.239 $\pm$ 0.106 & 1.055 $\pm$ 0.012 & 1.175 $\pm$ 0.028 & 1.376 $\pm$ 0.049 & 1.683 $\pm$ 0.081  \\\cline{2-11}

&188 & 26.509 $\pm$ 0.212 & 1.084 $\pm$ 0.008 & 1.264 $\pm$ 0.020 & 1.570 $\pm$ 0.036 & 2.061 $\pm$ 0.064 & 1.047 $\pm$ 0.008 & 1.144 $\pm$ 0.018 & 1.301 $\pm$ 0.030 & 1.528 $\pm$ 0.048  \\\cline{2-11}

&194 & 26.943 $\pm$ 0.316 & 1.092 $\pm$ 0.013 & 1.291 $\pm$ 0.030 & 1.638 $\pm$ 0.057 & 2.210 $\pm$ 0.104 & 1.055 $\pm$ 0.012 & 1.172 $\pm$ 0.027 & 1.367 $\pm$ 0.048 & 1.662 $\pm$ 0.079  \\\cline{2-11}

&200 & 27.548 $\pm$ 0.306 & 1.093 $\pm$ 0.011 & 1.296 $\pm$ 0.026 & 1.651 $\pm$ 0.050 & 2.235 $\pm$ 0.092 & 1.057 $\pm$ 0.011 & 1.180 $\pm$ 0.024 & 1.384 $\pm$ 0.042 & 1.693 $\pm$ 0.070 \\\cline{2-11}

&206 & 27.928 $\pm$ 0.259 & 1.093 $\pm$ 0.009 & 1.295 $\pm$ 0.021 & 1.646 $\pm$ 0.040 & 2.223 $\pm$ 0.072 & 1.057 $\pm$ 0.008 & 1.180 $\pm$ 0.019 & 1.383 $\pm$ 0.033 & 1.689 $\pm$ 0.055  \\\hline

& 91& 21.461 $\pm$ 0.107 & 1.092 $\pm$ 0.006 & 1.297 $\pm$ 0.015 & 1.670 $\pm$ 0.028 & 2.322 $\pm$ 0.051 & 1.045 $\pm$ 0.006 & 1.149 $\pm$ 0.013 & 1.333 $\pm$ 0.022 & 1.637 $\pm$ 0.037  \\\cline{2-11}

OPAL&161 & 24.378 $\pm$ 0.396 & 1.096 $\pm$ 0.020 & 1.310 $\pm$ 0.048 & 1.696 $\pm$ 0.093 & 2.361 $\pm$ 0.171 & 1.055 $\pm$ 0.020 & 1.179 $\pm$ 0.044 & 1.393 $\pm$ 0.077 & 1.739 $\pm$ 0.127  \\\cline{2-11}

&183 & 26.957 $\pm$ 0.272 & 1.101 $\pm$ 0.012 & 1.329 $\pm$ 0.028 & 1.750 $\pm$ 0.056 & 2.492 $\pm$ 0.107 & 1.064 $\pm$ 0.011 & 1.209 $\pm$ 0.026 & 1.470 $\pm$ 0.047 & 1.904 $\pm$ 0.082  \\\cline{2-11}

& 189& 27.007 $\pm$ 0.202 & 1.100 $\pm$ 0.009 & 1.319 $\pm$ 0.021 & 1.710 $\pm$ 0.040 & 2.377 $\pm$ 0.075 &1.063 $\pm$ 0.009 & 1.199 $\pm$ 0.019 & 1.434 $\pm$ 0.034 & 1.804 $\pm$ 0.057  \\\hline

\hline
\end{tabular}
\end{adjustbox}
\caption{Moments of Experimental, shifted Gompertz and Modified shifted Gompertz distributions for $e^{+}e^{-}$ collisions}
}
\end{table*}


\begin{table*}[t]
{\
\begin{adjustbox}{max width=\textwidth}

\begin{tabular}{|c|c|c|c|c|c|c|c|c|c|c|}
\hline
&  & &\multicolumn{4}{c|}{}&\multicolumn{4}{c|}{} \\

Energy& & &\multicolumn{4}{c|}{Normalized moments (Experiment)}&\multicolumn{4}{c|}{Normalized factorial moments (Experiment) } \\             

(GeV) & $|\eta|$&$<n>$ &\multicolumn{4}{c|}{}&\multicolumn{4}{c|}{}  \\\cline{4-11}

($p\overline{p}$) & & & & & & & & & &\\
& &  & C$_2$ & C$_3$ & C$_4$ & C$_5$ & F$_2$ & F$_3$ & F$_4$ & F$_5$                 \\\hline

& 0.5 & 2.501 $\pm$ 0.074 & 1.911 $\pm$ 0.053 & 4.895 $\pm$ 0.322 & 15.711 $\pm$ 1.876 & 60.718 $\pm$ 11.403 &1.511 $\pm$ 0.046 & 2.923 $\pm$ 0.238 & 6.945 $\pm$ 1.147 & 19.796 $\pm$ 5.362 \\\cline{2-11}

& 1.5 &7.940 $\pm$ 0.206 & 1.556 $\pm$ 0.037 & 3.159 $\pm$ 0.167 & 7.792 $\pm$ 0.695 & 22.349 $\pm$ 3.004 &1.430 $\pm$ 0.035 & 2.603 $\pm$ 0.144 & 5.664 $\pm$ 0.548 & 14.140 $\pm$ 2.136 \\\cline{2-11}

200 & 3 &15.540 $\pm$ 0.294 & 1.368 $\pm$ 0.024 & 2.329 $\pm$ 0.082 & 4.630 $\pm$ 0.249 & 10.275 $\pm$ 0.754 &1.304 $\pm$ 0.024 & 2.073 $\pm$ 0.075 & 3.791 $\pm$ 0.209 & 7.615 $\pm$ 0.577 \\\cline{2-11}

& 5 & 20.285 $\pm$ 0.479 & 1.278 $\pm$ 0.028 & 1.948 $\pm$ 0.087 & 3.379 $\pm$ 0.233 & 6.442 $\pm$ 0.605 & 1.229 $\pm$ 0.027 & 1.764 $\pm$ 0.080 & 2.836 $\pm$ 0.198 & 4.935 $\pm$ 0.470 \\\cline{2-11}

& Full &21.175 $\pm$ 0.500 & 1.244 $\pm$ 0.028 & 1.823 $\pm$ 0.082 & 3.023 $\pm$ 0.207 & 5.496 $\pm$ 0.510 & 1.197 $\pm$ 0.027 & 1.651 $\pm$ 0.075 & 2.536 $\pm$ 0.176 & 4.204 $\pm$ 0.395 \\\hline

& 0.5 & 3.004 $\pm$ 0.047 & 1.915 $\pm$ 0.029 & 5.037 $\pm$ 0.175 & 16.684 $\pm$ 1.004 & 66.035 $\pm$ 6.085 & 1.582 $\pm$ 0.026 & 3.347 $\pm$ 0.134 & 8.737 $\pm$ 0.654 & 26.799 $\pm$ 3.240 \\\cline{2-11}

& 1.5 &  9.471 $\pm$ 0.137 & 1.610 $\pm$ 0.022 & 3.480 $\pm$ 0.107 & 9.345 $\pm$ 0.482 & 29.750 $\pm$ 2.289 & 1.504 $\pm$ 0.021 & 2.992 $\pm$ 0.096 & 7.331 $\pm$ 0.403 & 21.149 $\pm$ 1.773 \\\cline{2-11}

540 & 3 & 18.968 $\pm$ 0.252 & 1.474 $\pm$ 0.017 & 2.857 $\pm$ 0.073 & 6.815 $\pm$ 0.284 & 19.083 $\pm$ 1.127 & 1.422 $\pm$ 0.017 & 2.630 $\pm$ 0.068 & 5.955 $\pm$ 0.254 & 15.758 $\pm$ 0.956 \\\cline{2-11}

& 5 & 26.330 $\pm$ 0.342 & 1.353 $\pm$ 0.016 & 2.271 $\pm$ 0.056 & 4.480 $\pm$ 0.176 & 10.016 $\pm$ 0.556 & 1.315 $\pm$ 0.015 & 2.120 $\pm$ 0.053 & 3.984 $\pm$ 0.159 & 8.425 $\pm$ 0.479 \\\cline{2-11}

& Full &28.375 $\pm$ 0.374 & 1.305 $\pm$ 0.016 & 2.079 $\pm$ 0.054 & 3.868 $\pm$ 0.159 & 8.128 $\pm$ 0.465 & 1.269 $\pm$ 0.016 & 1.943 $\pm$ 0.051 & 3.446 $\pm$ 0.143 & 6.853 $\pm$ 0.400 \\\hline

& 0.5 & 3.614 $\pm$ 0.087 & 1.940 $\pm$ 0.046 & 5.410 $\pm$ 0.318 & 19.789 $\pm$ 2.173 & 89.311 $\pm$ 15.604 & 1.664 $\pm$ 0.042 & 3.952 $\pm$ 0.269 & 12.314 $\pm$ 1.648 & 47.134 $\pm$ 10.245 \\\cline{2-11}

& 1.5 & 11.503 $\pm$ 0.247 & 1.582 $\pm$ 0.031 & 3.368 $\pm$ 0.144 & 8.736 $\pm$ 0.614 & 26.012 $\pm$ 2.634 & 1.496 $\pm$ 0.029 & 2.970 $\pm$ 0.131 & 7.107 $\pm$ 0.522 & 19.258 $\pm$ 2.066\\\cline{2-11}

900 & 3 &21.925 $\pm$ 0.428 & 1.456 $\pm$ 0.027 & 2.695 $\pm$ 0.103 & 5.828 $\pm$ 0.341 & 13.956 $\pm$ 1.108 &1.410 $\pm$ 0.026 & 2.500 $\pm$ 0.096 & 5.124 $\pm$ 0.303 & 11.487 $\pm$ 0.920 \\\cline{2-11}

& 5 &32.263 $\pm$ 0.642 & 1.354 $\pm$ 0.025 & 2.278 $\pm$ 0.088 & 4.464 $\pm$ 0.266 & 9.736 $\pm$ 0.797 & 1.323 $\pm$ 0.025 & 2.154 $\pm$ 0.083 & 4.054 $\pm$ 0.244 & 8.427 $\pm$ 0.698 \\\cline{2-11}

& Full & 35.632 $\pm$ 0.686 & 1.284 $\pm$ 0.023 & 2.003 $\pm$ 0.073 & 3.606 $\pm$ 0.203 & 7.185 $\pm$ 0.554 & 1.256 $\pm$ 0.022 & 1.897 $\pm$ 0.069 & 3.279 $\pm$ 0.186 & 6.227 $\pm$ 0.485 \\\hline

\hline
&  & &\multicolumn{4}{c|}{}&\multicolumn{4}{c|}{} \\

Energy& & &\multicolumn{4}{c|}{Normalized moments (shifted Gompertz)}&\multicolumn{4}{c|}{Normalized factorial moments (shifted Gompertz) } \\             

(GeV) & $|\eta|$&$<n>$ &\multicolumn{4}{c|}{}&\multicolumn{4}{c|}{}  \\\cline{4-11}

($p\overline{p}$) & & & & & & & & & &\\
& &  & C$_2$ & C$_3$ & C$_4$ & C$_5$ & F$_2$ & F$_3$ & F$_4$ & F$_5$                 \\\hline


& 0.5 & 2.475 $\pm$ 0.029 & 1.912 $\pm$ 0.024 & 4.864 $\pm$ 0.122 & 15.015 $\pm$ 0.573 & 53.249 $\pm$ 2.762 & 1.508 $\pm$ 0.020 & 2.873 $\pm$ 0.078 & 6.261 $\pm$ 0.264 & 14.708 $\pm$ 0.873 \\\cline{2-11}

& 1.5 & 7.917 $\pm$ 0.055 & 1.536 $\pm$ 0.011 & 3.018 $\pm$ 0.045 & 7.021 $\pm$ 0.158 & 18.455 $\pm$ 0.560 &1.410 $\pm$ 0.011 & 2.468 $\pm$ 0.037 & 4.992 $\pm$ 0.115 & 11.123 $\pm$ 0.349 \\\cline{2-11}

200 & 3 & 15.173 $\pm$ 0.061 & 1.381 $\pm$ 0.007 & 2.348 $\pm$ 0.022 & 4.626 $\pm$ 0.063 & 10.137 $\pm$ 0.181 &  1.315 $\pm$ 0.007 & 2.083 $\pm$ 0.020 & 3.762 $\pm$ 0.052 & 7.425 $\pm$ 0.134 \\\cline{2-11}

& 5 & 20.152 $\pm$ 0.094 & 1.297 $\pm$ 0.006 & 2.014 $\pm$ 0.020 & 3.560 $\pm$ 0.053 & 6.915 $\pm$ 0.137 & 1.247 $\pm$ 0.006 & 1.826 $\pm$ 0.018 & 2.995 $\pm$ 0.045 & 5.314 $\pm$ 0.106 \\\cline{2-11}

& Full &21.064 $\pm$ 0.152 & 1.245 $\pm$ 0.010 & 1.827 $\pm$ 0.028 & 3.039 $\pm$ 0.070 & 5.560 $\pm$ 0.171 & 1.198 $\pm$ 0.009 & 1.654 $\pm$ 0.026 & 2.549 $\pm$ 0.059 & 4.254 $\pm$ 0.132 \\\hline

& 0.5 & 2.976 $\pm$ 0.016 & 1.917 $\pm$ 0.012 & 5.082 $\pm$ 0.064 & 17.035 $\pm$ 0.324 & 67.968 $\pm$ 1.738 & 1.581 $\pm$ 0.010 & 3.376 $\pm$ 0.045 & 8.940 $\pm$ 0.182 & 27.480 $\pm$ 0.756 \\\cline{2-11}

& 1.5 & 9.428 $\pm$ 0.044 & 1.599 $\pm$ 0.008 & 3.421 $\pm$ 0.035 & 9.095 $\pm$ 0.143 & 28.618 $\pm$ 0.611 &  1.493 $\pm$ 0.008 & 2.935 $\pm$ 0.031 & 7.109 $\pm$ 0.115 & 20.226 $\pm$ 0.444 \\\cline{2-11}

540 & 3 &18.774 $\pm$ 0.076 & 1.448 $\pm$ 0.006 & 2.705 $\pm$ 0.023 & 6.155 $\pm$ 0.081 & 16.359 $\pm$ 0.293 & 1.394 $\pm$ 0.006 & 2.479 $\pm$ 0.021 & 5.335 $\pm$ 0.071 & 13.338 $\pm$ 0.243 \\\cline{2-11}

& 5 &26.154 $\pm$ 0.093 & 1.345 $\pm$ 0.005 & 2.258 $\pm$ 0.017 & 4.523 $\pm$ 0.053 & 10.440 $\pm$ 0.166 & 1.307 $\pm$ 0.005 & 2.107 $\pm$ 0.016 & 4.026 $\pm$ 0.048 & 8.822 $\pm$ 0.141 \\\cline{2-11}

& Full &28.112 $\pm$ 0.084 & 1.307 $\pm$ 0.004 & 2.099 $\pm$ 0.014 & 3.982 $\pm$ 0.040 & 8.625 $\pm$ 0.115 & 1.271 $\pm$ 0.004 & 1.962 $\pm$ 0.013 & 3.551 $\pm$ 0.036 & 7.299 $\pm$ 0.099 \\\hline

& 0.5 &3.555 $\pm$ 0.028 & 1.894 $\pm$ 0.016 & 4.905 $\pm$ 0.087 & 15.756 $\pm$ 0.424 & 59.066 $\pm$ 2.159 & 1.613 $\pm$ 0.015 & 3.465 $\pm$ 0.065 & 8.991 $\pm$ 0.259 & 26.371 $\pm$ 1.043 \\\cline{2-11}

& 1.5 &11.352 $\pm$ 0.054 & 1.557 $\pm$ 0.008 & 3.187 $\pm$ 0.032 & 7.850 $\pm$ 0.121 & 22.057 $\pm$ 0.460 & 1.469 $\pm$ 0.007 & 2.791 $\pm$ 0.029 & 6.294 $\pm$ 0.099 & 15.956 $\pm$ 0.342 \\\cline{2-11}

900 & 3 &21.835 $\pm$ 0.075 & 1.441 $\pm$ 0.005 & 2.604 $\pm$ 0.019 & 5.486 $\pm$ 0.060 & 12.867 $\pm$ 0.189 & 1.395 $\pm$ 0.005 & 2.410 $\pm$ 0.018 & 4.803 $\pm$ 0.053 & 10.539 $\pm$ 0.156 \\\cline{2-11}

& 5 &32.073 $\pm$ 0.101 & 1.324 $\pm$ 0.004 & 2.134 $\pm$ 0.014 & 3.980 $\pm$ 0.040 & 8.275 $\pm$ 0.111 & 1.293 $\pm$ 0.004 & 2.013 $\pm$ 0.013 & 3.595 $\pm$ 0.036 & 7.105 $\pm$ 0.096 \\\cline{2-11}

& Full &35.567 $\pm$ 0.157 & 1.268 $\pm$ 0.006 & 1.935 $\pm$ 0.018 & 3.406 $\pm$ 0.048 & 6.680 $\pm$ 0.125 & 1.240 $\pm$ 0.006 & 1.830 $\pm$ 0.017 & 3.091 $\pm$ 0.043 & 5.774 $\pm$ 0.109 \\\hline

\hline
&  & &\multicolumn{4}{c|}{}&\multicolumn{4}{c|}{} \\

Energy& & &\multicolumn{4}{c|}{Normalized moments (Modified shifted Gompertz)}&\multicolumn{4}{c|}{Normalized factorial moments (Modified shifted Gompertz) } \\             

(GeV) & $|\eta|$&$<n>$ &\multicolumn{4}{c|}{}&\multicolumn{4}{c|}{}  \\\cline{4-11}

($p\overline{p}$) & & & & & & & & & &\\
&  &  & C$_2$ & C$_3$ & C$_4$ & C$_5$ & F$_2$ & F$_3$ & F$_4$ & F$_5$                 \\\hline


& 0.5 & 2.477 $\pm$ 0.041 & 1.881 $\pm$ 0.033 & 4.604 $\pm$ 0.166 & 13.493 $\pm$ 0.742 & 45.166 $\pm$ 3.376 & 1.478 $\pm$ 0.028 & 2.651 $\pm$ 0.104 & 5.320 $\pm$ 0.326 & 11.404 $\pm$ 0.993 \\\cline{2-11}

& 1.5 & 7.895 $\pm$ 0.071 & 1.547 $\pm$ 0.016 & 3.085 $\pm$ 0.063 & 7.340 $\pm$ 0.226 & 19.818 $\pm$ 0.819 & 1.420 $\pm$ 0.015 & 2.529 $\pm$ 0.053 & 5.257 $\pm$ 0.167 & 12.101 $\pm$ 0.520 \\\cline{2-11}

200 & 3 & 15.537 $\pm$ 0.107 & 1.367 $\pm$ 0.010 & 2.321 $\pm$ 0.034 & 4.593 $\pm$ 0.101 & 10.130 $\pm$ 0.300 & 1.302 $\pm$ 0.009 & 2.065 $\pm$ 0.030 & 3.758 $\pm$ 0.084 & 7.493 $\pm$ 0.226 \\\cline{2-11}

& 5 &20.484 $\pm$ 0.134 & 1.270 $\pm$ 0.009 & 1.923 $\pm$ 0.026 & 3.317 $\pm$ 0.068 & 6.293 $\pm$ 0.172 & 1.221 $\pm$ 0.008 & 1.742 $\pm$ 0.024 & 2.786 $\pm$ 0.057 & 4.827 $\pm$ 0.134 \\\cline{2-11}

& Full &21.124 $\pm$ 0.185 & 1.245 $\pm$ 0.012 & 1.830 $\pm$ 0.036 & 3.054 $\pm$ 0.089 & 5.610 $\pm$ 0.216 & 1.197 $\pm$ 0.012 & 1.657 $\pm$ 0.033 & 2.565 $\pm$ 0.075 & 4.301 $\pm$ 0.167 \\\hline

& 0.5 &2.981 $\pm$ 0.022 & 1.901 $\pm$ 0.016 & 4.929 $\pm$ 0.085 & 15.986 $\pm$ 0.419 & 61.449 $\pm$ 2.190 & 1.566 $\pm$ 0.014 & 3.241 $\pm$ 0.059 & 8.192 $\pm$ 0.234 & 23.951 $\pm$ 0.945 \\\cline{2-11}

& 1.5 &9.463 $\pm$ 0.055 & 1.606 $\pm$ 0.011 & 3.468 $\pm$ 0.047 & 9.313 $\pm$ 0.192 & 29.587 $\pm$ 0.828 & 1.501 $\pm$ 0.010 & 2.981 $\pm$ 0.041 & 7.305 $\pm$ 0.154 & 21.009 $\pm$ 0.607 \\\cline{2-11}

540 & 3 &18.863 $\pm$ 0.092 & 1.470 $\pm$ 0.008 & 2.836 $\pm$ 0.031 & 6.714 $\pm$ 0.112 & 18.569 $\pm$ 0.421 &1.417 $\pm$ 0.007 & 2.608 $\pm$ 0.028 & 5.857 $\pm$ 0.099 & 15.278 $\pm$ 0.353 \\\cline{2-11} 
 
& 5 &26.179 $\pm$ 0.113 & 1.352 $\pm$ 0.006 & 2.282 $\pm$ 0.022 & 4.582 $\pm$ 0.067 & 10.568 $\pm$ 0.207 & 1.314 $\pm$ 0.006 & 2.130 $\pm$ 0.021 & 4.080 $\pm$ 0.060 & 8.931 $\pm$ 0.177 \\\cline{2-11}

& Full &28.144 $\pm$ 0.100 & 1.306 $\pm$ 0.005 & 2.096 $\pm$ 0.017 & 3.964 $\pm$ 0.048 & 8.553 $\pm$ 0.138 & 1.271 $\pm$ 0.005 & 1.959 $\pm$ 0.016 & 3.535 $\pm$ 0.043 & 7.234 $\pm$ 0.118 \\\hline

& 0.5 &3.578 $\pm$ 0.038 & 1.907 $\pm$ 0.022 & 5.041 $\pm$ 0.120 & 16.619 $\pm$ 0.631 & 63.995 $\pm$ 3.464 & 1.627 $\pm$ 0.019 & 3.599 $\pm$ 0.093 & 9.671 $\pm$ 0.412 & 29.394 $\pm$ 1.868 \\\cline{2-11}

& 1.5 &11.428 $\pm$ 0.071 & 1.575 $\pm$ 0.010 & 3.303 $\pm$ 0.044 & 8.345 $\pm$ 0.172 & 23.917 $\pm$ 0.675 & 1.487 $\pm$ 0.010 & 2.905 $\pm$ 0.039 & 6.740 $\pm$ 0.142 & 17.448 $\pm$ 0.511 \\\cline{2-11}

900& 3 &22.026 $\pm$ 0.098 & 1.449 $\pm$ 0.007 & 2.672 $\pm$ 0.026 & 5.755 $\pm$ 0.084 & 13.741 $\pm$ 0.266 & 1.404 $\pm$ 0.007 & 2.479 $\pm$ 0.024 & 5.060 $\pm$ 0.074 & 11.315 $\pm$ 0.221 \\\cline{2-11}

& 5 &32.341 $\pm$ 0.129 & 1.350 $\pm$ 0.006 & 2.259 $\pm$ 0.019 & 4.393 $\pm$ 0.056 & 9.482 $\pm$ 0.163 & 1.319 $\pm$ 0.006 & 2.136 $\pm$ 0.018 & 3.988 $\pm$ 0.052 & 8.197 $\pm$ 0.142 \\\cline{2-11}

& Full &35.734 $\pm$ 0.199 & 1.285 $\pm$ 0.008 & 2.015 $\pm$ 0.024 & 3.653 $\pm$ 0.066 & 7.354 $\pm$ 0.177 & 1.257 $\pm$ 0.008 & 1.908 $\pm$ 0.023 & 3.326 $\pm$ 0.060 & 6.386 $\pm$ 0.155 \\\hline 

\end{tabular}
\end{adjustbox}
\caption{Moments of Experimental, shifted Gompertz and modified shifted Gompertz distributions for $p\overline{p}$ collisions}
}
\end{table*}


\begin{table*}[t]
{\
\begin{adjustbox}{max width=\textwidth}
\begin{tabular}{|c|c|c|c|c|c|c|c|c|c|c|}
\hline
&  & &\multicolumn{4}{c|}{}&\multicolumn{4}{c|}{} \\

Energy& & &\multicolumn{4}{c|}{Normalized moments (Experiment)}&\multicolumn{4}{c|}{Normalized factorial moments (Experiment) } \\             

(GeV) & $|\eta|$&$<n>$ &\multicolumn{4}{c|}{}&\multicolumn{4}{c|}{}  \\\cline{4-11}

($pp$) & & & & & & & & & &\\
& &  & C$_2$ & C$_3$ & C$_4$ & C$_5$ & F$_2$ & F$_3$ & F$_4$ & F$_5$ \\\hline


&0.5 &4.426 $\pm$ 0.084 & 1.582 $\pm$ 0.036 & 3.386 $\pm$ 0.152 & 8.893 $\pm$ 0.592 & 27.013 $\pm$ 2.387 & 1.356 $\pm$ 0.032 & 2.416 $\pm$ 0.113 & 5.122 $\pm$ 0.356 & 12.119 $\pm$ 1.125 \\\cline{2-11}

&1.0 &8.059 $\pm$ 0.121 & 1.570 $\pm$ 0.029 & 3.272 $\pm$ 0.117 & 8.229 $\pm$ 0.432 & 23.642 $\pm$ 1.636 & 1.446 $\pm$ 0.027 & 2.719 $\pm$ 0.099 & 6.047 $\pm$ 0.324 & 15.050 $\pm$ 1.064 \\\cline{2-11}

900&1.5 &11.744 $\pm$ 0.151 & 1.538 $\pm$ 0.024 & 3.092 $\pm$ 0.094 & 7.413 $\pm$ 0.331 & 20.116 $\pm$ 1.188 & 1.453 $\pm$ 0.023 & 2.714 $\pm$ 0.083 & 5.952 $\pm$ 0.269 & 14.542 $\pm$ 0.874 \\\cline{2-11}

&2.0 &15.507 $\pm$ 0.183 & 1.504 $\pm$ 0.022 & 2.912 $\pm$ 0.080 & 6.659 $\pm$ 0.270 & 17.095 $\pm$ 0.914 & 1.440 $\pm$ 0.021 & 2.630 $\pm$ 0.073 & 5.599 $\pm$ 0.229 & 13.205 $\pm$ 0.714  \\\cline{2-11}

&2.4 &18.432 $\pm$ 0.211 & 1.474 $\pm$ 0.020 & 2.759 $\pm$ 0.073 & 6.042 $\pm$ 0.235 & 14.748 $\pm$ 0.755 & 1.420 $\pm$ 0.020 & 2.525 $\pm$ 0.067 & 5.190 $\pm$ 0.203 & 11.742 $\pm$ 0.607 \\\hline

&0.5 &5.289 $\pm$ 0.095 & 1.585 $\pm$ 0.034 & 3.334 $\pm$ 0.138 & 8.443 $\pm$ 0.521 & 24.490 $\pm$ 2.041 & 1.396 $\pm$ 0.031 & 2.507 $\pm$ 0.107 & 5.243 $\pm$ 0.337 & 12.194 $\pm$ 1.086 \\\cline{2-11}

&1.0 &9.722 $\pm$ 0.144 & 1.565 $\pm$ 0.027 & 3.185 $\pm$ 0.106 & 7.689 $\pm$ 0.383 & 21.099 $\pm$ 1.426 & 1.463 $\pm$ 0.026 & 2.723 $\pm$ 0.092 & 5.900 $\pm$ 0.300 & 14.287 $\pm$ 0.998 \\\cline{2-11}

2360&1.5 &13.968 $\pm$ 0.196 & 1.523 $\pm$ 0.025 & 2.956 $\pm$ 0.093 & 6.738 $\pm$ 0.316 & 17.392 $\pm$ 1.114 & 1.451 $\pm$ 0.024 & 2.640 $\pm$ 0.084 & 5.552 $\pm$ 0.264 & 13.072 $\pm$ 0.858 \\\cline{2-11}

&2.0 &18.323 $\pm$ 0.251 & 1.490 $\pm$ 0.023 & 2.806 $\pm$ 0.085 & 6.177 $\pm$ 0.282 & 15.347 $\pm$ 0.955 & 1.435 $\pm$ 0.023 & 2.568 $\pm$ 0.079 & 5.306 $\pm$ 0.245 & 12.256 $\pm$ 0.776 \\\cline{2-11}

&2.4 &21.696 $\pm$ 0.298 & 1.471 $\pm$ 0.023 & 2.736 $\pm$ 0.084 & 6.008 $\pm$ 0.280 & 15.092 $\pm$ 0.973 & 1.425 $\pm$ 0.022 & 2.537 $\pm$ 0.078 & 5.285 $\pm$ 0.249 & 12.518 $\pm$ 0.823  \\\hline

&0.5 &6.924 $\pm$ 0.077 & 1.732 $\pm$ 0.026 & 4.159 $\pm$ 0.114 & 12.245 $\pm$ 0.486 & 41.444 $\pm$ 2.156 &  1.588 $\pm$ 0.024 & 3.450 $\pm$ 0.097 & 9.020 $\pm$ 0.366 & 26.546 $\pm$ 1.413 \\\cline{2-11}

&1.0 &13.122 $\pm$ 0.118 & 1.705 $\pm$ 0.021 & 3.945 $\pm$ 0.088 & 11.020 $\pm$ 0.353 & 35.044 $\pm$ 1.461 & 1.629 $\pm$ 0.020 & 3.567 $\pm$ 0.081 & 9.323 $\pm$ 0.301 & 27.410 $\pm$ 1.154 \\\cline{2-11}

7000&1.5 &19.421 $\pm$ 0.154 & 1.669 $\pm$ 0.018 & 3.725 $\pm$ 0.073 & 9.946 $\pm$ 0.279 & 30.027 $\pm$ 1.094 & 1.617 $\pm$ 0.018 & 3.472 $\pm$ 0.069 & 8.843 $\pm$ 0.249 & 25.240 $\pm$ 0.926 \\\cline{2-11}

&2.0 &25.870 $\pm$ 0.188 & 1.640 $\pm$ 0.016 & 3.562 $\pm$ 0.064 & 9.187 $\pm$ 0.234 & 26.573 $\pm$ 0.879 & 1.601 $\pm$ 0.016 & 3.375 $\pm$ 0.061 & 8.388 $\pm$ 0.215 & 23.203 $\pm$ 0.771 \\\cline{2-11}

&2.4&30.790 $\pm$ 0.213 & 1.616 $\pm$ 0.015 & 3.433 $\pm$ 0.059 & 8.613 $\pm$ 0.209 & 24.169 $\pm$ 0.760 &  1.584 $\pm$ 0.015 & 3.277 $\pm$ 0.056 & 7.963 $\pm$ 0.194 & 21.495 $\pm$ 0.678 \\\hline 

\hline
&  & &\multicolumn{4}{c|}{}&\multicolumn{4}{c|}{} \\

Energy& & &\multicolumn{4}{c|}{Normalized moments (shifted Gompertz)}&\multicolumn{4}{c|}{Normalized factorial moments (shifted Gompertz) } \\             

(GeV) & $|\eta|$&$<n>$ &\multicolumn{4}{c|}{}&\multicolumn{4}{c|}{}  \\\cline{4-11}

($pp$) & & & & & & & & & &\\
& &  & C$_2$ & C$_3$ & C$_4$ & C$_5$ & F$_2$ & F$_3$ & F$_4$ & F$_5$                 \\\hline
  
&0.5&4.433 $\pm$ 0.033 & 1.571 $\pm$ 0.017 & 3.348 $\pm$ 0.065 & 8.804 $\pm$ 0.245 & 26.927 $\pm$ 0.973 & 1.345 $\pm$ 0.015 & 2.387 $\pm$ 0.048 & 5.083 $\pm$ 0.146 & 12.190 $\pm$ 0.453 \\\cline{2-11}

&1.0&8.099 $\pm$ 0.038 & 1.522 $\pm$ 0.010 & 3.056 $\pm$ 0.038 & 7.474 $\pm$ 0.132 & 21.215 $\pm$ 0.482 & 1.399 $\pm$ 0.010 & 2.522 $\pm$ 0.032 & 5.455 $\pm$ 0.097 & 13.480 $\pm$ 0.310 \\\cline{2-11}

900&1.5&11.768 $\pm$ 0.038 & 1.521 $\pm$ 0.007 & 3.019 $\pm$ 0.025 & 7.213 $\pm$ 0.086 & 19.783 $\pm$ 0.307 & 1.437 $\pm$ 0.007 & 2.645 $\pm$ 0.022 & 5.791 $\pm$ 0.070 & 14.371 $\pm$ 0.226 \\\cline{2-11}

&2.0&15.536 $\pm$ 0.040 & 1.499 $\pm$ 0.005 & 2.887 $\pm$ 0.019 & 6.604 $\pm$ 0.062 & 17.122 $\pm$ 0.209 & 1.435 $\pm$ 0.005 & 2.606 $\pm$ 0.017 & 5.556 $\pm$ 0.052 & 13.271 $\pm$ 0.163 \\\cline{2-11}

&2.4&18.445 $\pm$ 0.043 & 1.470 $\pm$ 0.004 & 2.737 $\pm$ 0.016 & 5.996 $\pm$ 0.049 & 14.790 $\pm$ 0.159 & 1.416 $\pm$ 0.004 & 2.504 $\pm$ 0.014 & 5.152 $\pm$ 0.043 & 11.810 $\pm$ 0.128 \\\hline

&0.5&5.275 $\pm$ 0.035 & 1.580 $\pm$ 0.015 & 3.318 $\pm$ 0.056 & 8.439 $\pm$ 0.206 & 24.846 $\pm$ 0.798 & 1.390 $\pm$ 0.013 & 2.491 $\pm$ 0.043 & 5.249 $\pm$ 0.132 & 12.514 $\pm$ 0.422 \\\cline{2-11}

&1.0&9.608 $\pm$ 0.040 & 1.570 $\pm$ 0.009 & 3.206 $\pm$ 0.034 & 7.845 $\pm$ 0.119 & 22.148 $\pm$ 0.447 & 1.466 $\pm$ 0.008 & 2.738 $\pm$ 0.029 & 6.023 $\pm$ 0.093 & 15.112 $\pm$ 0.314 \\\cline{2-11}

2360&1.5&13.762 $\pm$ 0.045 & 1.536 $\pm$ 0.007 & 3.013 $\pm$ 0.024 & 7.000 $\pm$ 0.082 & 18.718 $\pm$ 0.297 & 1.464 $\pm$ 0.006 & 2.688 $\pm$ 0.022 & 5.774 $\pm$ 0.069 & 14.160 $\pm$ 0.231 \\\cline{2-11}

&2.0&17.994 $\pm$ 0.051 & 1.512 $\pm$ 0.005 & 2.884 $\pm$ 0.020 & 6.472 $\pm$ 0.065 & 16.550 $\pm$ 0.224 & 1.456 $\pm$ 0.005 & 2.639 $\pm$ 0.018 & 5.560 $\pm$ 0.056 & 13.253 $\pm$ 0.183 \\\cline{2-11}

&2.4&21.374 $\pm$ 0.055 & 1.486 $\pm$ 0.005 & 2.795 $\pm$ 0.017 & 6.276 $\pm$ 0.058 & 16.423 $\pm$ 0.207 & 1.439 $\pm$ 0.005 & 2.591 $\pm$ 0.016 & 5.527 $\pm$ 0.052 & 13.694 $\pm$ 0.176 \\\hline

&0.5&7.178 $\pm$ 0.028 & 1.652 $\pm$ 0.010 & 3.749 $\pm$ 0.040 & 10.568 $\pm$ 0.158 & 34.861 $\pm$ 0.668 & 1.513 $\pm$ 0.009 & 3.098 $\pm$ 0.034 & 7.771 $\pm$ 0.118 & 22.472 $\pm$ 0.435 \\\cline{2-11}

&1.0&13.744 $\pm$ 0.032 & 1.615 $\pm$ 0.006 & 3.488 $\pm$ 0.022 & 9.202 $\pm$ 0.082 & 28.179 $\pm$ 0.321 & 1.542 $\pm$ 0.006 & 3.146 $\pm$ 0.020 & 7.771 $\pm$ 0.070 & 22.099 $\pm$ 0.253 \\\cline{2-11}

7000&1.5&20.066 $\pm$ 0.034 & 1.609 $\pm$ 0.004 & 3.426 $\pm$ 0.016 & 8.811 $\pm$ 0.056 & 26.028 $\pm$ 0.213 & 1.560 $\pm$ 0.004 & 3.191 $\pm$ 0.015 & 7.829 $\pm$ 0.050 & 21.925 $\pm$ 0.180 \\\cline{2-11}

&2.0&25.909 $\pm$ 0.034 & 1.642 $\pm$ 0.003 & 3.559 $\pm$ 0.012 & 9.216 $\pm$ 0.045 & 27.029 $\pm$ 0.169 & 1.604 $\pm$ 0.003 & 3.371 $\pm$ 0.012 & 8.419 $\pm$ 0.041 & 23.653 $\pm$ 0.148 \\\cline{2-11}

&2.4&30.658 $\pm$ 0.034 & 1.628 $\pm$ 0.003 & 3.471 $\pm$ 0.010 & 8.794 $\pm$ 0.037 & 25.165 $\pm$ 0.136 & 1.596 $\pm$ 0.003 & 3.313 $\pm$ 0.010 & 8.134 $\pm$ 0.034 & 22.423 $\pm$ 0.122 \\\hline

\hline
&  & &\multicolumn{4}{c|}{}&\multicolumn{4}{c|}{} \\

Energy& & &\multicolumn{4}{c|}{Normalized moments (Modified shifted Gompertz)}&\multicolumn{4}{c|}{Normalized factorial moments (Modified shifted Gompertz) } \\             

(GeV) & $|\eta|$&$<n>$ &\multicolumn{4}{c|}{}&\multicolumn{4}{c|}{}  \\\cline{4-11}

($pp$) & & & && & & & & &\\
&  &  & C$_2$ & C$_3$ & C$_4$ & C$_5$ & F$_2$ & F$_3$ & F$_4$ & F$_5$                 \\\hline

&0.5&4.423 $\pm$ 0.046 & 1.579 $\pm$ 0.024 & 3.376 $\pm$ 0.094 & 8.878 $\pm$ 0.352 & 27.104 $\pm$ 1.379 &  1.353 $\pm$ 0.022 & 2.407 $\pm$ 0.069 & 5.117 $\pm$ 0.207 & 12.219 $\pm$ 0.634 \\\cline{2-11}

&1.0&8.097 $\pm$ 0.048 & 1.534 $\pm$ 0.014 & 3.101 $\pm$ 0.049 & 7.601 $\pm$ 0.171 & 21.522 $\pm$ 0.618 & 1.411 $\pm$ 0.013 & 2.563 $\pm$ 0.042 & 5.549 $\pm$ 0.126 & 13.651 $\pm$ 0.395 \\\cline{2-11}

900&1.5&11.784 $\pm$ 0.051 & 1.521 $\pm$ 0.010 & 3.020 $\pm$ 0.035 & 7.199 $\pm$ 0.117 & 19.628 $\pm$ 0.410 & 1.436 $\pm$ 0.009 & 2.647 $\pm$ 0.031 & 5.778 $\pm$ 0.095 & 14.234 $\pm$ 0.300 \\\cline{2-11}

&2.0&15.510 $\pm$ 0.054 & 1.497 $\pm$ 0.008 & 2.891 $\pm$ 0.027 & 6.624 $\pm$ 0.087 & 17.169 $\pm$ 0.290 &  1.433 $\pm$ 0.007 & 2.609 $\pm$ 0.024 & 5.573 $\pm$ 0.074 & 13.299 $\pm$ 0.226 \\\cline{2-11}

&2.4&18.433 $\pm$ 0.059 & 1.465 $\pm$ 0.007 & 2.732 $\pm$ 0.023 & 5.995 $\pm$ 0.072 & 14.792 $\pm$ 0.228 & 1.411 $\pm$ 0.007 & 2.500 $\pm$ 0.021 & 5.152 $\pm$ 0.062 & 11.809 $\pm$ 0.183 \\\hline

&0.5&5.309 $\pm$ 0.050 & 1.575 $\pm$ 0.021 & 3.290 $\pm$ 0.080 & 8.292 $\pm$ 0.292 & 24.062 $\pm$ 1.111 & 1.386 $\pm$ 0.019 & 2.471 $\pm$ 0.062 & 5.148 $\pm$ 0.186 & 12.032 $\pm$ 0.579 \\\cline{2-11}

&1.0&9.746 $\pm$ 0.052 & 1.554 $\pm$ 0.012 & 3.141 $\pm$ 0.044 & 7.570 $\pm$ 0.151 & 20.928 $\pm$ 0.546 & 1.452 $\pm$ 0.011 & 2.684 $\pm$ 0.038 & 5.810 $\pm$ 0.117 & 14.237 $\pm$ 0.380 \\\cline{2-11}

2360&1.5&13.978 $\pm$ 0.062 & 1.516 $\pm$ 0.009 & 2.937 $\pm$ 0.033 & 6.711 $\pm$ 0.108 & 17.510 $\pm$ 0.374 & 1.445 $\pm$ 0.009 & 2.622 $\pm$ 0.030 & 5.533 $\pm$ 0.090 & 13.208 $\pm$ 0.288 \\\cline{2-11}

&2.0&18.284 $\pm$ 0.069 & 1.486 $\pm$ 0.008 & 2.797 $\pm$ 0.027 & 6.183 $\pm$ 0.085 & 15.515 $\pm$ 0.283 & 1.431 $\pm$ 0.007 & 2.559 $\pm$ 0.025 & 5.313 $\pm$ 0.074 & 12.414 $\pm$ 0.230 \\\cline{2-11}

&2.4&21.760 $\pm$ 0.079 & 1.463 $\pm$ 0.007 & 2.710 $\pm$ 0.024 & 5.949 $\pm$ 0.078 & 15.039 $\pm$ 0.263 & 1.417 $\pm$ 0.007 & 2.513 $\pm$ 0.022 & 5.235 $\pm$ 0.069 & 12.498 $\pm$ 0.221 \\\hline

&0.5&6.979 $\pm$ 0.035 & 1.707 $\pm$ 0.013 & 4.024 $\pm$ 0.056 & 11.645 $\pm$ 0.226 & 38.944 $\pm$ 0.966 & 1.564 $\pm$ 0.013 & 3.331 $\pm$ 0.047 & 8.554 $\pm$ 0.168 & 24.908 $\pm$ 0.623 \\\cline{2-11}

&1.0&13.220 $\pm$ 0.041 & 1.677 $\pm$ 0.008 & 3.800 $\pm$ 0.033 & 10.424 $\pm$ 0.125 & 32.771 $\pm$ 0.502 & 1.601 $\pm$ 0.008 & 3.431 $\pm$ 0.030 & 8.802 $\pm$ 0.106 & 25.612 $\pm$ 0.394 \\\cline{2-11}

7000&1.5&19.665 $\pm$ 0.047 & 1.634 $\pm$ 0.006 & 3.561 $\pm$ 0.024 & 9.316 $\pm$ 0.086 & 27.750 $\pm$ 0.324 & 1.583 $\pm$ 0.006 & 3.317 $\pm$ 0.022 & 8.275 $\pm$ 0.076 & 23.324 $\pm$ 0.273 \\\cline{2-11}

&2.0&25.817 $\pm$ 0.048 & 1.636 $\pm$ 0.005 & 3.547 $\pm$ 0.018 & 9.152 $\pm$ 0.065 & 26.588 $\pm$ 0.240 & 1.597 $\pm$ 0.005 & 3.360 $\pm$ 0.017 & 8.354 $\pm$ 0.059 & 23.225 $\pm$ 0.210 \\\cline{2-11}

&2.4&30.731 $\pm$ 0.050 & 1.612 $\pm$ 0.004 & 3.417 $\pm$ 0.015 & 8.582 $\pm$ 0.054 & 24.209 $\pm$ 0.193 &1.579 $\pm$ 0.004 & 3.262 $\pm$ 0.015 & 7.934 $\pm$ 0.050 & 21.540 $\pm$ 0.172 \\\hline 

\end{tabular}
\end{adjustbox}
\caption{Moments of Experimental, shifted Gompertz and Modified shifted Gompertz distributions for $pp$ collisions}
}
\end{table*}

\begin{table*}[t]
\centering

{
 \begin{adjustbox}{max width=\textwidth}

\begin{tabular}{|c|c|c|c|c|c|c|c|c|c|}
\hline
  & &\multicolumn{4}{c|}{}&\multicolumn{4}{c|}{} \\

 Energy& &\multicolumn{4}{c|}{Normalized moments (Predicted)}&\multicolumn{4}{c|}{Normalized factorial moments (Predicted) } \\             

(GeV)&$<n>$ &\multicolumn{4}{c|}{}&\multicolumn{4}{c|}{}  \\\cline{3-10}

($e^{+}e^{-}$) & & & & & & & & &\\
 &  & C$_2$ & C$_3$ & C$_4$ & C$_5$ & F$_2$ & F$_3$ & F$_4$ & F$_5$                 \\\hline
&&&&&&&&&\\
 500  & 37.081 $\pm$ 0.087& 1.145 $\pm$ 0.003 & 1.487 $\pm$ 0.008 & 2.167 $\pm$ 0.018 & 3.496 $\pm$ 0.038 & 1.118 $\pm$ 0.003 & 1.396 $\pm$ 0.008 & 1.936 $\pm$ 0.016 & 2.948 $\pm$ 0.032  \\
&&&&&&&&&\\\hline

\end{tabular}
\end{adjustbox}
\caption{Predicted moments at $\sqrt{s}$ = 500 $GeV$ from shifted Gompertz distribution for $e^{+}e^{-}$ collisions}
}
\end{table*}
    
\section{Uncertainties on Moments}

Given a distribution $P(n)$ which is normalized to unity with an uncertainty $\epsilon_{n}$, and assuming that the errors on the individual bins are uncorrelated, the moment errors can be calculated by using the method described in \cite{Diso}, using the partial derivatives:

\begin{equation} 
\frac{\partial C_q}{\partial P_n} = \frac{n^{q}\langle n \rangle - \langle n^{q} \rangle qn}{\langle n \rangle ^{q+1}}
\end{equation}
\begin{equation}
\frac{\partial F_q}{\partial P_n} = \frac{n(n-1)....(n-q+1)\langle n \rangle - \langle n(n-1)......(n-q+1)\rangle qn}{\langle n \rangle ^{q+1}} 
\end{equation}
The total error is then
\begin{equation}
E_{q}^2 = \sum_{n}\Big(\frac {\partial X_{q}}{\partial P_n}\epsilon _n \Big)^2 
\end{equation} 
where $X_{q}$ is $C_{q}$ or $F_{q}$.\\

In the published data, multiplicities are given either as (value + statistical error + systematic error) or as (value + error).~The error on the multiplicities have been taken as the total error, by adding the statistical and systematic errors in quadrature.

\section{Conclusion}

An analysis of moments of multiplicity distributions described within a newly proposed statistical distribution, the shifted Gompertz distribution and its modified form has been done.~We had proposed and shown that the use of this statistical distribution for studying the multiplicity distributions in high energy collisions reproduces the results in $e^{+}e^{-}$, $p\overline{p}$ and $pp$ collisions very well.

A good agreement between the normalized moments as well as normalized factorial moments obtained from the shifted Gompertz distribution and its modified form with the experimental values, serves as a good test of the validity of the proposed distribution.~The results have reproduced the violation of KNO scaling as observed for higher moments in the measured data.~At higher energies, particularly at LHC energies, the moments strongly are dependent on the energy.~The information dissemination from such an analysis is often used to study the patterns and correlations in the multi-particle final state of high-energy collisions in the presence of statistical fluctuations.~In this connection we find that the factorial moments are large indicating correlations amongst the produced particles.~The predictions for normalized moments and normalized factorial moments are also made for $e^+e^-$ collisions at 500 GeV at a future collider. 

\section{Data Availability}
All the data used in the paper can be obtained from the references quoted or from the authors.
\section*{References}

\bibliography{mybibfile}

\end{document}